\documentclass[twocolumn]{aastex631}

\shorttitle{Modelling energetic proton transport during GLE events}
\shortauthors{Waterfall et al.}
\graphicspath{{./}{figures/}}

\begin{document}

\title{Modelling the transport of relativistic solar protons along a heliospheric current sheet during historic GLE events}

\author[0000-0003-4390-2920]{Charlotte O.G. Waterfall}
\affiliation{Jeremiah Horrocks Institute \\
University of Central Lancashire \\
PR1 2HE, UK}

\author[0000-0002-7837-5780]{Silvia Dalla}
\affiliation{Jeremiah Horrocks Institute \\
University of Central Lancashire \\
PR1 2HE, UK}

\author[0000-0002-7719-7783]{Timo Laitinen}
\affiliation{Jeremiah Horrocks Institute \\
University of Central Lancashire \\
PR1 2HE, UK}

\author{Adam Hutchinson}
\affiliation{Jeremiah Horrocks Institute \\
University of Central Lancashire \\
PR1 2HE, UK}

\author[0000-0003-2765-0874]{Mike Marsh}
\affiliation{Met Office \\
Exeter \\
UK}

\begin{abstract}
There are many difficulties associated with forecasting high-energy solar particle events at Earth. One issue is understanding why some large solar eruptive events trigger ground level enhancement (GLE) events and others do not. In this work we perform 3D test particle simulations of a set of historic GLEs to understand more about what causes these powerful events. Particular focus is given to studying how the heliospheric current sheet (HCS) affects high-energy proton transport through the heliosphere following an event. Analysis of $\geq$M7.0 flares between 1976$-$2020 shows that active regions located closer to the HCS ($<$10$^{\circ}$) are more likely to be associated with a GLE event. We found that modelled GLE events where the source region was close to the HCS also led to increased heliospheric transport in longitude and higher count rates (when the Earth was located in the drift direction). In a model that does not include perpendicular diffusion associated with turbulence, the HCS is the dominant mechanism affecting heliospheric particle transport for GLE 42 and 69, and varying other parameters (e.g. a narrow, 10$^{\circ}$, or wider, 60$^{\circ}$, injection width) causes little change. Overall in our model, the HCS is relevant in 71$\%$ of our analysed GLEs and including it more accurately reproduces observed intensities near Earth. Our simulations enable us to produce model profiles at Earth that can be compared to existing observations by the GOES satellites and neutron monitors, as well as for use in developing future forecasting models. 
\end{abstract}


\section{Introduction}\label{sec.intro}
 
The most powerful solar eruptive events, such as large flares and coronal mass ejections (CMEs), are capable of accelerating particles to relativistic energies. These particles propagate through interplanetary (IP) space where they can be detected in the near-Earth environment. A fraction of these solar energetic particle (SEP) events comprise ions of energies up to tens of GeV. Signatures of these relativistic particles are observed on the surface of Earth in what is known as ground level enhancement (GLE) events (\cite{shea2012space}, \cite{nitta2012special}). Whilst events as powerful as these are more rare, they are important to include in space weather forecasting models due to the risk they pose to life and infrastructure in space (\citet{siscoe2000space}, \citet{schwenn2006space}).

There have been 72 GLE events recorded between 1942 and 2017 \citep{usoskin2020revised}. The first GLEs were detected through ionisation chambers \citep{forbush1946three}, eventually being replaced by a network of neutron monitors across the globe. GLE occurrence is more likely around solar maximum, when large flares and fast CMEs are more common. There have been few space-based instruments capable of recording fluxes up to relativistic energies, however the High Energy Proton and Alpha Detector (HEPAD) onboard the Geostationary Operational Environmental Satellites (GOES) have recorded proton fluxes up to 700\,MeV since the 1980s \citep{onsager1996operational}. GOES/HEPAD has observed numerous relativistic proton events so far, most of which did not register as GLEs. Of the five largest GLEs recorded since 1956 \citep{mccracken2012high}, three have been observed by HEPAD onboard GOES as well as neutron monitors. One of these GLEs, GLE 69 on 20 January 2005, with neutron monitor increases orders of magnitude larger than its counterparts, has been the subject of much research. For example, \cite{mccracken2008investigation} evaluated observations from numerous sources in detail and discussed two possible explanations for the particle acceleration during GLE 69: driven by the flare itself or by shocks associated with CMEs. Indeed, the origin of this acceleration for relativistic particles is often debated, with CME driven shocks frequently suggested to be the preferential cause of the acceleration (e.g. \citet{reames2009solar}, \citet{gopalswamy2013first}, \citet{cliver2020solar}). \citet{grechnev2008extreme} studied the associated flare and emissions from GLE 69 and suggested the flare, not the CME, was responsible for this huge event. 

Forecasting SEP events is difficult for many reasons. There are many uncertainties related to the location and size of the source region of the particles. The nature of the acceleration of the particles (e.g. flare driven or from a CME shock) as well as the propagation through IP space (e.g. the degree of transport across the magnetic field) also have large uncertainties associated with them. The behaviour and transport of the particles through IP space is a particularly important factor in these computationally demanding models. While a source location near the heliolongitudinal range of 40$-$60 on the western hemisphere is favourable due to the Parker spiral magnetic connectivity with the Earth, the turbulent conditions of the heliospheric plasma can dramatically influence the transport of the particles to 1AU. Forecasting GLE and high energy SEP events is particularly challenging due to the relative lack of both frequency of these events as well as comprehensive data sets on them. 

To understand the complexity behind the GLEs, case studies of transport in individual GLEs have been performed (e.g. GLE 60 \citep{bieber2004spaceship}, GLE 69 \citep{saiz2005relativistic}). Additionally, modelling of energetic particle propagation has highlighted the importance of particle drifts during SEP events
(e.g. \citet{dalla2013solar}, \citet{dalla2015drift}). The latter of these research areas has recently involved exploring the effect the heliospheric current sheet (HCS) has on SEP particle propagation (\citet{kubo20092}, \citet{battarbee2017solar}).  The heliospheric current sheet delimits the boundary between the solar dipole field's hemispheres and has been seen to affect the drift direction of energetic particles longitudinally under a variety of conditions \citep{battarbee2018modeling}. Previous test particle models of particle propagation in the heliosphere have included those with a flat current sheet \citep{battarbee2017solar} and a range of wavy configurations \citep{battarbee2018modeling}. \citet{dalla20203d} investigated the propagation of relativistic particles for a range of monoenergetic particle populations and scattering conditions, and showed that the HCS allows efficient propagation in the direction perpendicular to the average magnetic field.
Perpendicular transport may also be caused by turbulence effects, in particular by random walk of the magnetic field lines, as discussed in several recent studies (e.g. \citet{tooprakai2016simulations}, \citet{laitinen2016solar}, \citet{laitinen2018effect}.)

In this work we focus on evaluating the role of the HCS on the propagation of high energy particles. In particular, we assess whether the current sheet configuration present during a selection of historic GLE events has a large influence on proton propagation to 1AU and hence whether it is necessary to include in future forecasting tools. We investigate this through 3D test particle modelling of multiple historic events, as well as statistical analysis of source regions relative to the HCS between 1976 and 2020. Previous simulations of these events have mainly been limited to 1D models or single GLEs, e.g. \citet{bieber2002energetic}. \citet{augusto2018relativistic} analysed data for the 2017 September 10 GLE (GLE 72) and proposed that propagation along the HCS was the main mechanism allowing relativistic protons to reach Earth. They also studied an additional 7 GLE or sub-GLE events and stated that in all cases the source active region was located within an HCS structure. In 6 of these, Earth was also located within it.  In this paper however, we are modelling multiple historic events with a 3D test particle code that includes a HCS configuration unique to each event. This way, the simulated intensity profiles can be compared to those observed by HEPAD and neutron monitors. We are also specifically modelling the transport of $>$300\,MeV protons within the heliosphere, which experience faster travel and stronger drifts across the heliosphere.

The selection of GLE events chosen is outlined in Section \ref{sec.events}, with a discussion of the test particle model and relevant parameters in Section \ref{sec.simulations}. Section \ref{sec.analysisgle} discusses some initial analysis of all historic GLE events. Section \ref{sec.data_analysis} details the results of our simulations and includes comparison with HEPAD and neutron monitor data. We simulate our GLEs both with and without a current sheet to determine its influence in specific events. Section \ref{sec.conclusions} presents our conclusions and discussion of heliospheric transport in historic GLE events. 

\section{Selection of historic events} 
\label{sec.events}
The selection of the GLE events for our study is based on several criteria. As the simulations will be compared to observed profiles from GOES/HEPAD observations we are limited to GLE events occurring between 1984 to 2017. This reduces the full set of 72 GLEs to 34. We also limit our sample to GLEs with polar neutron monitor increases of 15$\%$ or higher (as noted in \citet{mccracken2012high} and the revised GLE database described in \citet{usoskin2020revised}). This reduces the sample to 17 GLEs. GLE 47, 48 and 59 were removed as their current sheet configuration was too complex for our current model to fit (see Section \ref{subsec.fitting}).


\begin{deluxetable*}{cccccccc}
\tablenum{1}
\tablecaption{List of GLE events, ordered by date, modelled with the 3D test particle code. GLEs are selected according to criteria in Section \ref{sec.events}. \label{tab.eventstab}}
\tablewidth{0pt}
\tablehead{\colhead{Event} & \colhead{Flare} & \colhead{Polarity} & \colhead{Vsw} & \colhead{$\Delta$f}  & \colhead{$\Delta$E} & \colhead{$\psi$} & \colhead{Counts at Earth}\\
\colhead{(GLE \#)} & \colhead{location} & \colhead{} & \colhead{/ km\,s$^{-1}$} &  \colhead{$^{\circ}$}  & \colhead{$^\circ$} & \colhead{$^\circ$} & \colhead{with no HCS?}
}
\decimalcolnumbers
\startdata
1984 Feb 02 (39) & S16W94 & $-$ & 500* & 8.9	& 15.7 & 26.4 & None\\ 
1989 Aug 16 (41) & S18W85 & $-$ & 500* & 18.0 & 6.8 &  31.3 & None \\
1989 Sep 29 (42) & S36W90 & $-$ & 500* & 7.2 &	5.1 &	35.6 & Yes$^{b}$ \\ 
1989 Oct 19 (43) & S27E10 & $-$ & 500* & 5.9 &	5.6	& 71.2 & None \\ 
1989 Oct 22 (44) & S27W31 & $-$ & 500* & 6.7 &	6.9	& 41.8 & None \\ 
1989 Oct 24 (45) & S30W57 & $-$ & 500* & 8.5 &	14.6 &	35.1 & None \\ 
1991 Jun 15 (52) & N33W70 & + & 500* &  49.3 &	32.8 &	34.9 & Yes$^{a}$\\ 
1997 Nov 06 (55) & S18W63 & + & 350 & 17.7 & 0.7 & 24.9 & Yes$^{b}$ \\ 
2001 Apr 15 (60) & S20W85 & $-$ & 480 & 6.3 &	11.8 &	31.4 & Yes$^{b}$ \\
2001 Apr 18 (61) & S20W117 & $-$ & 500 & 5.6 &  10.8 & 52.6 & None \\ 
2003 Oct 28 (65) & S16E08 & $-$ & 730 & 28.8 &	42.7 &	52.3 & Yes$^{a}$ \\
2003 Oct 29 (66) & S15W09 & $-$ & 1000 & 29.2 & 48.3 & 38.5 & Yes$^{a}$ \\ 
2003 Nov 02 (67) & S14W59 & $-$ & 510 & 30.5 & 48.9 & 18.4 & Yes$^{a}$ \\ 
2005 Jan 20 (69) & N14W63 & $-$ & 820 & 5.3 & 0.6	& 29.8 & None \\ 
2006 Dec 13 (70) & S06W24 & $-$ & 650 & 13.9 & 25.0 & 14.5 & Yes$^{a}$ \\ 
2012 May 17 (71) & S11W76 & $-$ & 350 & 13.9 & 14.8 & 14.7 & None \\ 
2017 Sep 10 (72) & S08W88 & + & 510 & 15.6 & 8.3 & 41.7 & None \\
\enddata
\tablecomments{The polarity of the IMF (A$-$ or A+) is given by + or $-$ in Column 3. The solar wind speed (km\,s$^{-1}$) used for each simulation is listed in Column 4 (* values denote those where solar wind data was unavailable so 500km\,s$^{-1}$ was chosen). Column 5, 6 and 7 give information on the angular distance between the flare and HCS ($\Delta$f), Earth's footpoint and HCS ($\Delta$E) and between the flare and Earth's footpoint ($\psi$) respectively. Column 8 highlights those events where there were no counts at Earth's location when the HCS was removed from the simulation. Events marked $^{a}$ and $^{b}$ in Column 8 indicate events where there were counts at Earth without a HCS, and the HCS was found to be irrelevant or relevant respectively. (see Section \ref{subsec.nohcs} for discussion).}
\end{deluxetable*}

The full list of events we have simulated is given in Table \ref{tab.eventstab}. There are 17 GLE events in total, however only the results for GLEs 42, 65 and 69 are discussed in detail in this paper. We highlight GLEs 42 and 69 as they are the largest GLEs (i.e. largest neutron monitor increases) in our sample, and learning more about what caused the magnitude of these events is critical for future forecasting models. GLE 42 and 69 are among the largest GLEs ever recorded, along with GLE 5. However, there is little data available from GLE 5 (1956) so we have been unable to model it here. Although GLE 69 has been extensively documented in literature, the role of the current sheet through modelling for this event, and others, has not been addressed before. The simulation of GLE 65 is also highlighted in this paper, as it is a good counterexample of an Eastern event where neither the flare or Earth is located near the current sheet but still leads to a GLE event.

Within our sample there are only 3 of 17 GLE events where the flare location and Earth's footpoint (at the solar surface, as determined by magnetic connection along the Parker Spiral magnetic field) have a small longitudinal separation (less than 20$^{\circ}$). The majority of events have a large longitudinal separation between the Earth and flare. This supports the idea that there is a large degree of longitudinal spread in the heliosphere during SEP events.

\begin{figure}[t]
\epsscale{1.2}
\centering
\plotone{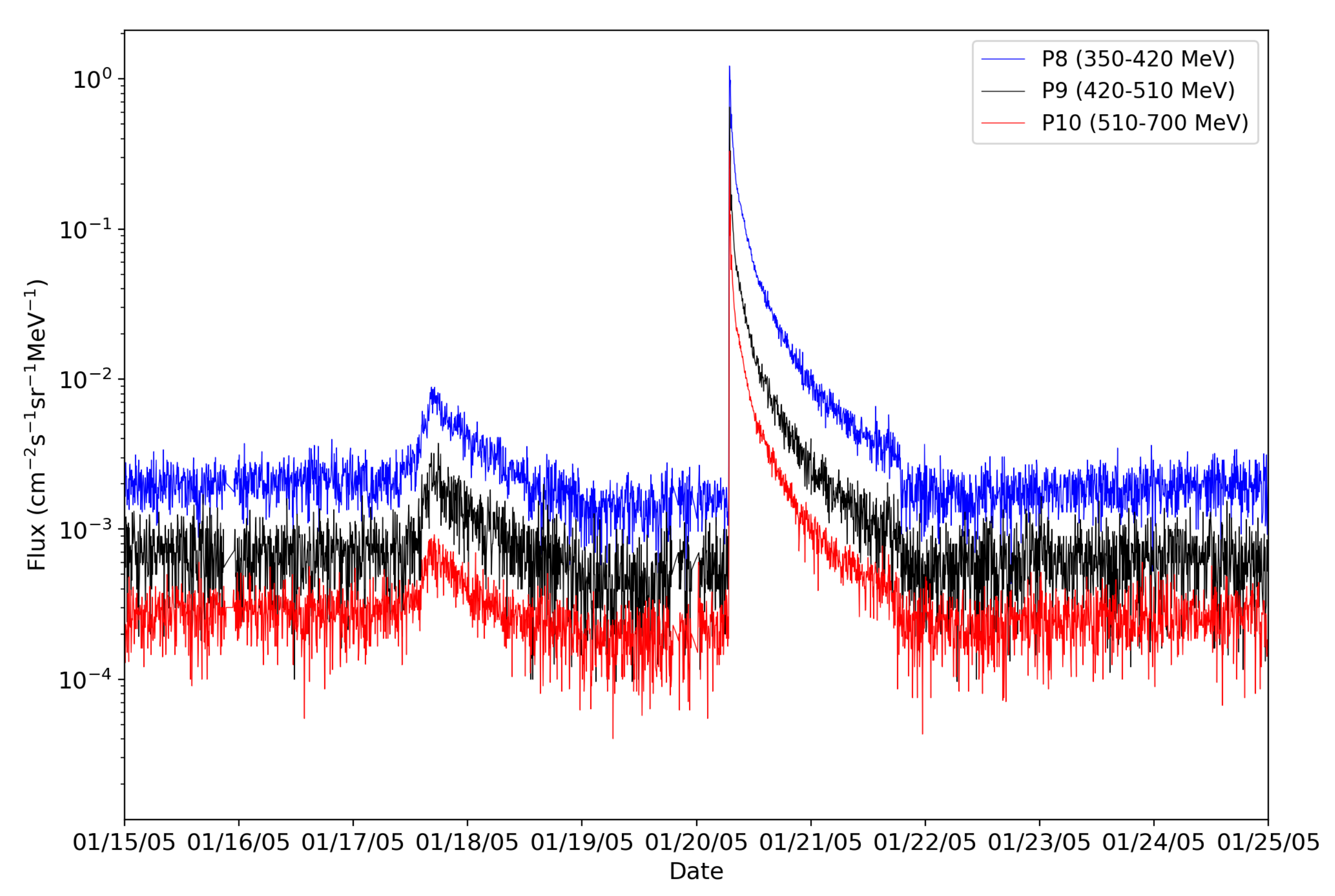}
\caption{GOES HEPAD intensity profiles in the P8, P9 and P10 (blue, black, red) high energy differential proton channels from 15$-$25th January 2005. GLE 68 and 69 are observed as the first and second events respectively occurring on the 17th and 20th January.} 
\label{fig.jan05hepad}
\end{figure}

The GOES HEPAD flux profiles for GLE 69 are shown in Figure \ref{fig.jan05hepad} for channels P8, P9 and P10 (350$-$420, 420$-$510, 510$-$700\,MeV respectively). These are the largest HEPAD fluxes reported for any event. These profiles will be used in Section \ref{sec.data_analysis} for comparison with our model profiles.

Table \ref{tab.eventstab} lists the flare location associated with each GLE (\citet{belov2010ground}, \citet{mccracken2012high}, \citet{papaioannou2016solar}). There is some discrepancy with these locations and other catalogs, for example GLE 42 has a longitude of W105 in \citet{mccracken2012high}. In those cases, the most frequently reported longitude value is used. 

We have not excluded any GLEs with flares that occurred behind-the-limb (GLE 39 and 42). Despite its behind-the-limb origin, GLE 42 was one of the largest observed GLE events and is discussed in detail in Section \ref{sec.data_analysis}.

\begin{figure*}[ht]
\epsscale{0.9}
\centering
\plotone{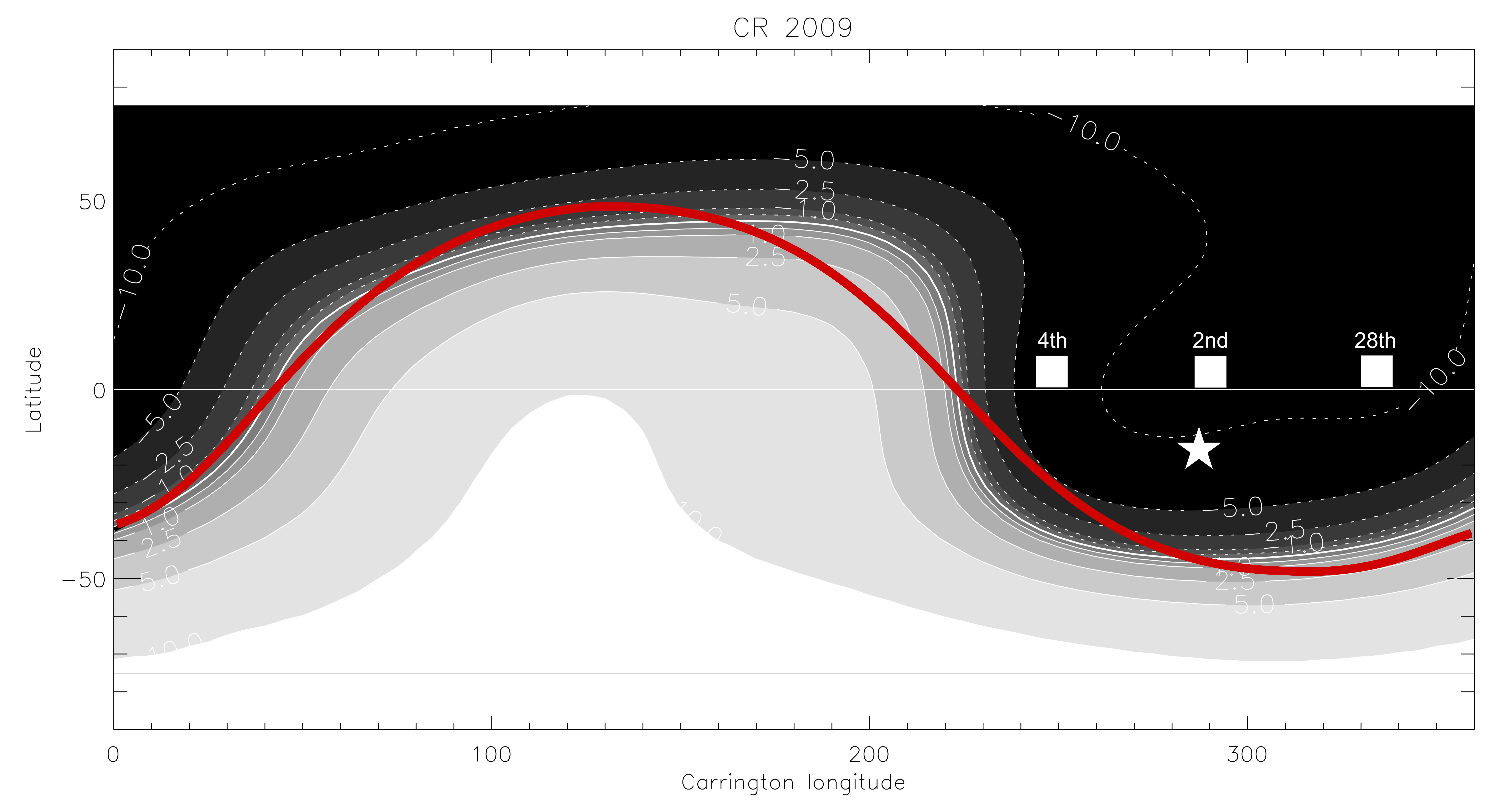}
\caption{Source surface map for Carrington rotation 2009 (2003-Oct-23 to 2003-Nov-19). The flare and Earth's relative footpoint (for three events, see Section  \ref{subsec.october}) are illustrated as the star and square symbols respectively. The dashed lines are the fit to the heliospheric current sheet described by the SSM's neutral line (solid white line). The neutral line separates the two hemispheres of oppositely directed magnetic field lines. The darker (lighter) region represents inward (outward) pointing field lines.} 
\label{fig.28octSSM}
\end{figure*}

We have ensured that our simulated GLEs cover a range of Earth footpoint and flare positions relative to the current sheet. Also, events that occur during different IMF configurations (either A+ or A$-$) are represented to see the effect on particle drift. The current sheet separates two hemispheres of oppositely directed open magnetic field lines. An A+ IMF polarity describes a configuration where the open magnetic field lines mostly point outwards in the northern hemisphere and inwards in the southern hemisphere. An A$-$ polarity describes the opposite field line configuration to A+, with field lines pointing outwards in the southern hemisphere and inwards in the north. The orientation is decided by examining the synoptic source surface maps provided by the Wilcox Solar Observatory (WSO) at: \url{http://wso.stanford.edu}. These source surface maps (SSMs) are produced from potential field modelling using photospheric magnetogram data. A SSM is produced for each Carrington rotation, so multiple events can sometimes be plotted on one map. Figure \ref{fig.28octSSM} shows an example SSM for Carrington rotation 2009. In this map, the darker region corresponds to inward pointing open magnetic field lines, and outward field lines for the lighter region. These two sectors represent the two hemispheres that are separated by the neutral line (the solid white line). In this case, the configuration is an A$-$. The HCS configuration used in our model is based off a fit to this neutral line. 

\begin{figure}[ht]
\epsscale{1.0}
\centering
\plotone{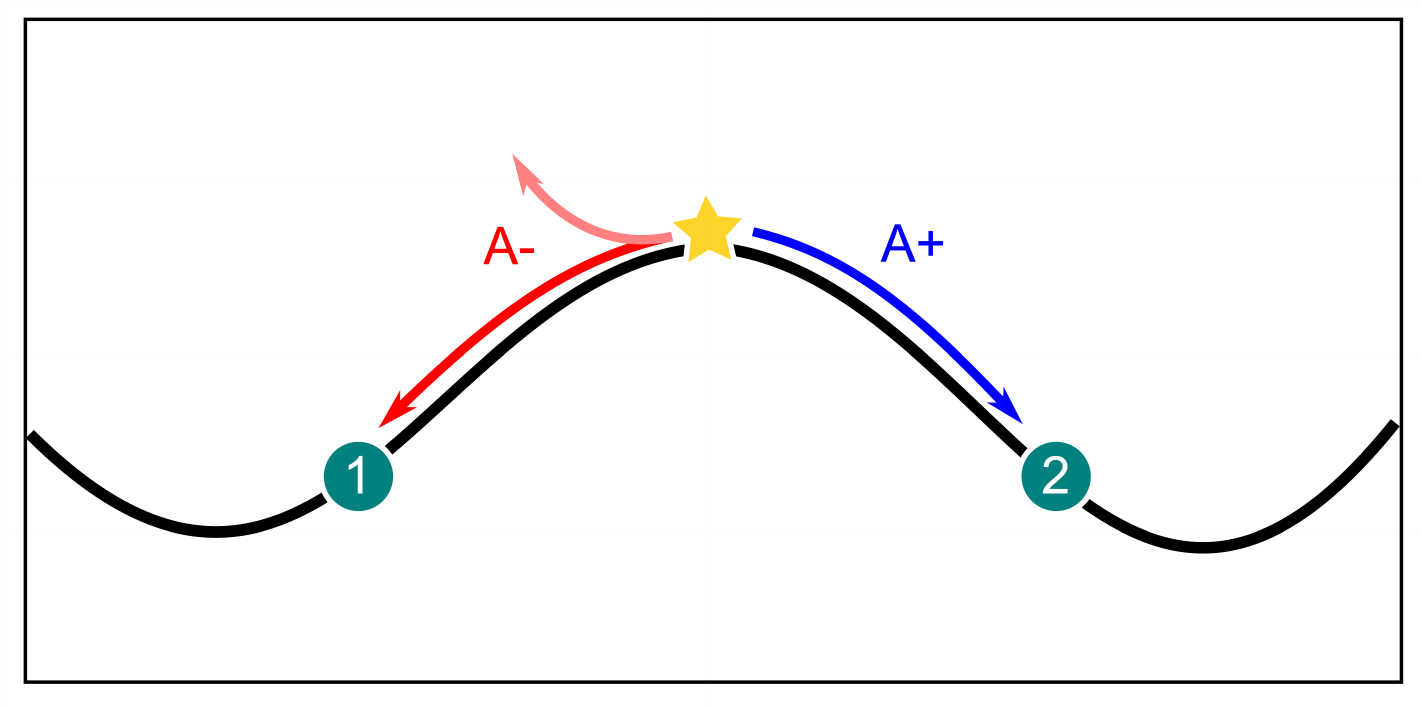}
\caption{Illustration showing particle drift directions depending on the IMF configuration. When the particles are injected close to the HCS (symbolised by the star), the particles drift eastwards (westwards) towards observer 1 (2) in an A$-$ (A+) configuration. There is also some drift away from the HCS in an A$-$ configuration.} 
\label{fig.hcsfigure}
\end{figure}

Different HCS configurations were explored by \citet{battarbee2018modeling}, who included current sheet configurations with different polarities to investigate the effect on particle propagation. They found that for an A+ IMF polarity the particle drift is focused towards and along the current sheet in a Westerly direction. For an A$-$ polarity, the drift is primarily towards the poles away from the current sheet. If the injection location is near to the current sheet in an A$-$ configuration, there is also drift towards the East along the sheet. This directional motion is illustrated in Figure \ref{fig.hcsfigure} using an example when the particles are injected directly onto a wavy HCS. In this paper we use East and West to refer to the directions as for locations on the Sun, e.g. an event on the Sun's Western limb occurs on the right hand edge to an observer at Earth (where solar north is upward). 

\section{Simulations}\label{sec.simulations}
The transport of relativistic protons during historic GLEs is simulated using a 3D test particle code first developed by \citet{dalla2005particle}. The model has been adapted to explore the behaviour of energetic protons within the heliosphere under various conditions. For example, \citet{kelly2012cross} included a Parker spiral field to investigate particle trajectories, \citet{battarbee2017solar} explored the transport of SEPs near a flat current sheet. The model used here includes drift effects from the curvature and gradient of the Parker spiral as well as HCS drifts. 
We include the effect of turbulence modelled as pitch angle scattering of protons as they propagate through the heliosphere. This is accounted for in our simulations using a Poisson distribution of scattering intervals for each particle, with a mean scattering time = $\lambda$/$v$ where $v$ is the particle velocity. The mean free path, $\lambda$, controls the degree of scattering present in the model. In line with the analysis of previous GLE events (\citet{bieber2002energetic}, \citet{bieber2004spaceship}, \citet{saiz2005relativistic}) we use $\lambda$\,=\,0.3AU as standard in all our simulations, but vary it between 0.1 and 0.5AU to represent different scattering conditions in the heliosphere. More detail on this simulated scattering and how its value affects heliospheric transport at high energies is given by \citet{dalla20203d}. 

While the scattering does cause some motion of the protons across the field, we have not explicitly included a term describing turbulence induced perpendicular transport (e.g. from field line meandering, \citet{laitinen2016solar}). We acknowledge that perpendicular transport is a potentially important factor in SEP events, however it is non-trivial to include in our current model and we stress that our primary focus is that of HCS induced transport relative to the polarity (drift direction) and proximity of the source. There are also large uncertainties in the parameters associated with perpendicular transport between events \citep{kocharov2020interplanetary}, and exploring this for each separate GLE is beyond the scope of this paper. It is even more uncertain how perpendicular effects such as the random walk of field lines behave close to the HCS. Despite this, perpendicular transport is likely to help in transporting the particles towards the HCS and hence towards the Earth and we hope to include it in a future development of our model.

The test particle model has many other parameters that can be altered depending on the event. The following subsections lists the parameters of interest in these simulations and, where necessary, justification for their values.

\subsection{Test particle model parameters}
\label{subsec.params}
Each GLE event was simulated using 3 million protons which were allowed to propagate for a total of 72 hours. This length of time was long enough to generate full intensity time profiles longer than any observed GLE period without exhausting too much computation time. While the test particle model is capable of simulating a variety of particles, we focused on the propagation of high energy protons here. As we are modelling GLE events, which are a product of high energy particles capable of penetrating the Earth's atmosphere, and to enable comparison with HEPAD data we chose to model protons in the energy range 300\,$\leq$\,E\,$\leq$\,1200\,MeV. We are simulating a CME-like injection, where the protons are injected instantaneously (not as a time-extended injection) according to a power law in energy with a spectral index, $\gamma$, that varies between $-$1.5 and $-$5. These were chosen to explore the effects of different values on the simulated intensity profiles. We use $-$1.5 as our `standard' value, representative of indices reported for SEP events derived from SOHO-EPHIN observations \citep{kuhl2017solar}.

The protons are injected instantaneously at a height of 2\,R$_\odot$ from the centre of the Sun \citep{kahler1994injection}. We have varied the injection region between a 10x10$^{\circ}$ and 60x60$^{\circ}$ region. We have chosen this lower value to represent the suggestion that the observed longitudinal extent of GLE events is smaller than that of SEP events at lower energies \citep{nitta2012special}. This may be taken as indication that only the nose of the shock is able to accelerate particles to relativistic energies \citep{hu2017modeling}.


We take the CME-like particle injection to be instantaneous and do not consider time-extended injection of particles by, e.g., a moving shock front. The acceleration of the particles themselves is not modelled. For each GLE, we take the coordinates of the associated flare as the source location, and use these coordinates to set the location of the injection region. 

The solar wind speed is individually set for each simulation corresponding to 5-minute averaged measurements taken by the CELIAS/MTOF Proton Monitor on-board the SOHO Spacecraft around the time of the event. In some cases, e.g. for GLE 39, solar wind data is unavailable so the standard value of 500 km\,s$^{-1}$ is chosen. This value is chosen based on solar wind speeds during other GLEs. The solar wind speed values for our modelled GLEs are shown in Table \ref{tab.eventstab} with the smallest speeds being no less than 350 km\,s$^{-1}$. 

Some parameters are varied in our simulations however, unless stated otherwise, the `standard' value of them used is: mean free path\,=\,0.3\,AU, injection width\,=\,10x10$^{\circ}$, power law index, $\gamma$\,=\,1.5. The magnetic field value is constant, scaled to be 3.85nT at 1AU, as in \cite{battarbee2018modeling}.

\subsection{Fitting the heliospheric current sheet}
\label{subsec.fitting}
As discussed in Section \ref{sec.events}, we use source surface maps (produced from potential field modelling performed by WSO) to obtain our HCS configuration. We model the configuration of the HCS based off of fits to the neutral line in a SSM (solid white line in Figure \ref{fig.28octSSM}). Figure \ref{fig.28octSSM} shows a sinusoidal fit to the neutral line, or current sheet, that is incorporated into the test particle model. This model fit, shown as the dashed line, can be altered depending on the behaviour of the current sheet during the Carrington rotation. The SSMs are available at radial source surface's of 2.5R$_\odot$ and 3.25R$_\odot$. In this paper we have only used the maps generated at a source surface of 2.5R$_\odot$. The SSM in Figure \ref{fig.28octSSM} is an example of a generally good fit in our model across the entire longitudinal range.  A more in-depth description of the current sheet and fitting procedure can be found in \citet{battarbee2018modeling}. In the model, the current sheet is assumed to have a thickness of 5000km at 1AU, as suggested by \citet{winterhalter1994heliospheric}. The IMF polarity is interchangeable within the model to be either A+, A$-$ or a unipolar field. For all of our simulated events we chose either an A+ or A$-$ configuration based upon its SSM orientation at the time of the GLE.  

\section{Analysis of historic GLE events}
\label{sec.analysisgle}
\begin{figure}[ht]
\epsscale{1.0}
\centering
\plotone{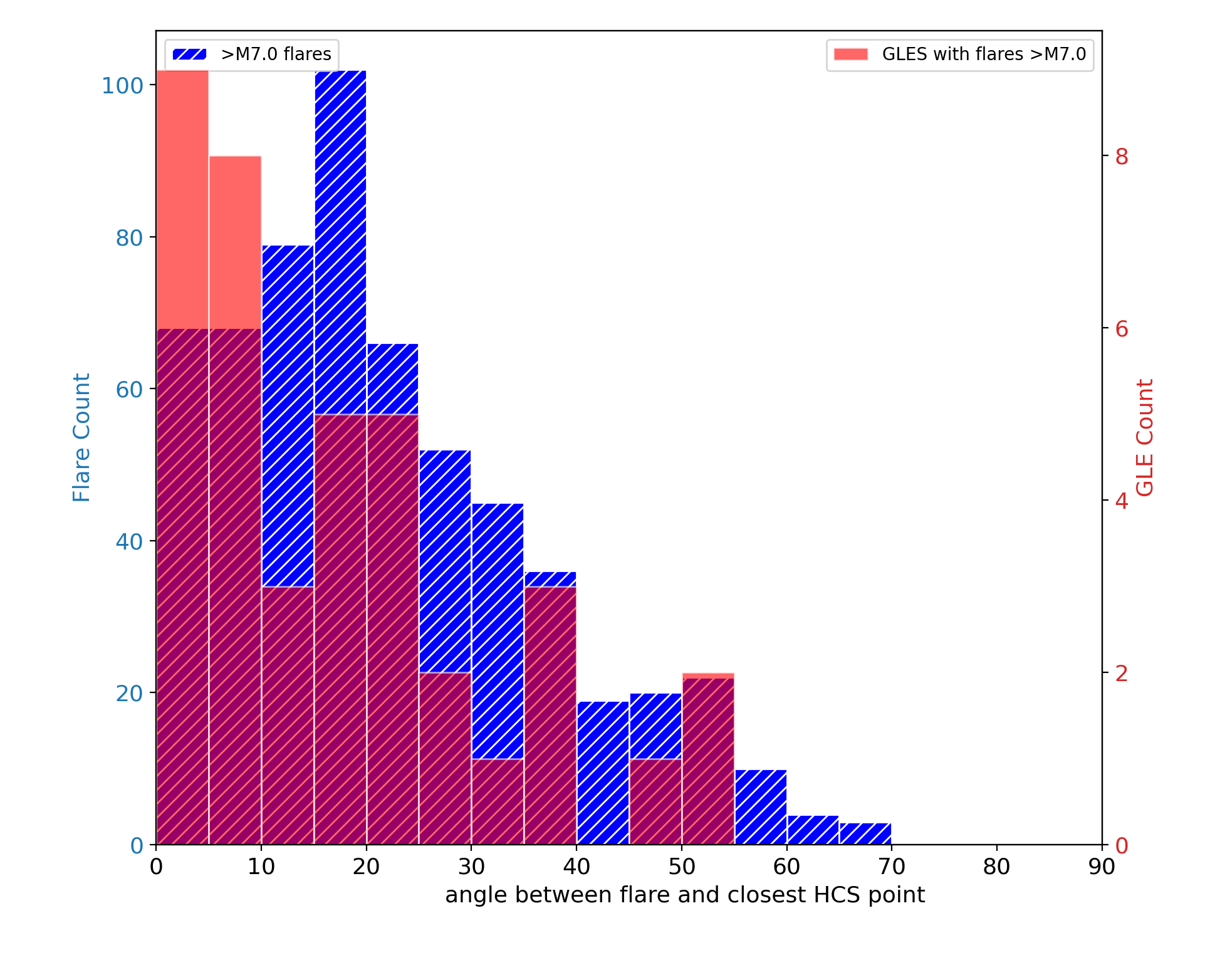}
\caption{Histogram of the angle subtended by the arc that joins the flare site and closest point on the HCS of all $\geq$M7.0 flares (blue) between 1976 and 2020. The flares associated with GLE events are shown in red.} 
\label{fig.histo}
\end{figure}

We have purposefully included events with large ($>$15$\%$) neutron monitor increases to see whether the heliospheric current sheet has any affect on these events (e.g. the October 1989 series). In Table \ref{tab.eventstab}, the events with an A$+$ polarity have the smallest neutron monitor increases. This could be due to the more focused transport of particles along the HCS (illustrated in Figure \ref{fig.hcsfigure}) during an A+ polarity, leading to less particles drifting towards Earth if it is not situated close to the HCS. In an A$-$ configuration there is some degree of drift away from the HCS, increased when the flare is further from the HCS. The relationship between the proximity of the flare to the HCS and the frequency of GLE occurrence is explored in Figure \ref{fig.histo}. 

Figure \ref{fig.histo} shows the distribution of all $\geq$M7.0 flare locations between 1976$-$2020 (blue) and for those flares associated with a GLE (red) binned according to their angular distance to the HCS. Note that the GLE sample size is considerably smaller than the flare sample. The data is taken between 1976$-$2020 (limited by the availability of GOES X-ray data) and is obtained from the Heliophysics Event Catalogue as part of HELIO (\url{helio-vo.eu}). As we are concerned with flares $\geq$M7.0 we have excluded GLEs 33, 35, 39, 40, 61, 71. Multiple events that occur from a single active region have been included.

The flares associated with GLEs are peaked closest to the HCS, with 44$\%$ occurring within 10$^{\circ}$ of the HCS. On the other hand, for non-GLE flares only 22$\%$ occur in this range and they peak in the 15$-$20$^{\circ}$ bin. The two-sample Kolmogorov-Smirnov test was performed and a p-value of 0.019 was obtained, suggesting the two samples are drawn from different distributions. This supports the idea that when a large flare occurs close to the HCS, it is more likely to produce a GLE event. This could be due to the HCS induced drifts, or from a possible more efficient production of relativistic particles for solar eruptive events near the HCS. This analysis suggests an important role of closeness to the HCS for GLE production, which will be explored further through simulations in the following section. 

As the Earth orbits within 7$^{\circ}$ of the solar equator, it is common for Earth to be close to the HCS (when the tilt angle of the HCS is not too large). However the position of its magnetic footpoint relative to the active region under different polarities can vary. Therefore it is useful to also look at the HCS drift direction for GLE events. 

For the 39 GLEs in Figure \ref{fig.histo}, 20 occur under an A$-$ configuration (19 for A+). As illustrated in Figure \ref{fig.hcsfigure}, for an A$-$ configuration the optimal position for the Earth's footpoint is towards the left of the flare (right for A+). For the GLEs in Figure \ref{fig.histo} where longitudinal transport is relevant (i.e. a flare and Earth footpoint separation of $>$20$^{\circ}$), 21/35 events have the optimal configuration for drifts. 86$\%$ of these events have neutron monitor percentage increases above 10$\%$. In contrast, of the 14 events where the Earth is on the `wrong side' of the flare, only 57$\%$ of them have percentage increases above 10$\%$.

When we examine which GLEs have their flares closest to the HCS we find that many of them have large neutron monitor increases also. The GLEs with the largest neutron monitor increases since 1984 are (with percentages from \citet{mccracken2012high}): GLE 69 (5500$\%$), GLE 42 (395$\%$), GLE 60 (230$\%$), GLE 44 (190$\%$), GLE 70 (110$\%$), GLE 39 (100$\%$), GLE 45 (95$\%$), GLE 43 (90$\%$), GLE 59 (80$\%$). Nearly all of these GLEs had flares less than 10$^{\circ}$ from the HCS, with GLE 69 and 42 having flares less than 5$^{\circ}$ from the HCS. The exception is GLE 70, which had a flare approximately 18$^{\circ}$ from the HCS, but the longitudinal separation between the Earth's footpoint and flare was minimal. This suggests that when a flare is close to the HCS, i.e. within 10$^{\circ}$, it can increase the severity of the event at Earth. 

This paper focuses on two of the largest GLEs on record; GLE 42 and GLE 69, which had flares very close to the HCS. They also have two of the three highest fluxes recorded from HEPAD P8, P9 and P10 channels. The other highest GOES flux is for the 14th July 2000, GLE 59, (neutron monitor increase of 80$\%$) which has been excluded from our current simulations because of its complex current sheet configuration, however it also had a flare within 10$^{\circ}$ of the HCS. GLE 42 and 69 have A$-$ configurations with the Earth's footpoint lying Eastward of the flare, directly in the path of the drift along the HCS from the injection region. Other GLEs with similarly strong flares and CMEs that do not have a flare or Earth footpoint located as close to the current sheet have neutron monitor increases 1-2 orders of magnitude smaller than GLE 42 and 69. 

However, not all GLEs with source regions close to the HCS have these large increases. GLE 56, with its small neutron monitor increase (less than 10$\%$) and close flare proximity to the HCS, is a good example of this. The IMF polarity at the time of the event was an A+, causing efficient drifts to the right along the HCS with minimal drift in other regions (see Figure \ref{fig.hcsfigure}). The Earth, despite being to the right of the flare, was situated North (and away from) the HCS during the event and was therefore out of the primary drift area leading to a reduced flux at Earth.  Simulations have been performed to confirm and understand more about how the HCS affects the particle propagation towards Earth from this event, and the others listed in Table \ref{tab.eventstab}. 

\section{Simulation results}\label{sec.data_analysis}
Our simulations enable us to analyse the extent to which the HCS aids in distributing particles away from the injection location throughout the heliosphere. In terms of particle counts at a specific observer, the ideal scenario for maximum counts appears to be where the flare and observer's footpoint are situated directly on the current sheet, with the observer on the `correct' side of the flare according to the polarity of the field (East for A$-$, West for A+) (see Figure \ref{fig.hcsfigure}).


The large GLEs, 42 and 69, have this ideal flare and Earth footpoint configuration. The results from the test particle simulations for these two large events are discussed here, with a complementary example for the 28 October 2003 (GLE 65) also shown. This event is a good comparison event where the flare and Earth are located far from the current sheet. GLE 65 was the first in a series of GLEs that were followed by the largest flare on record (4 November 2003, X28 flare, no GLE produced) which is also discussed here.

\subsection{GLE 42: 29 September 1989}
\begin{figure*}[t]
\epsscale{0.9}
\centering
\plotone{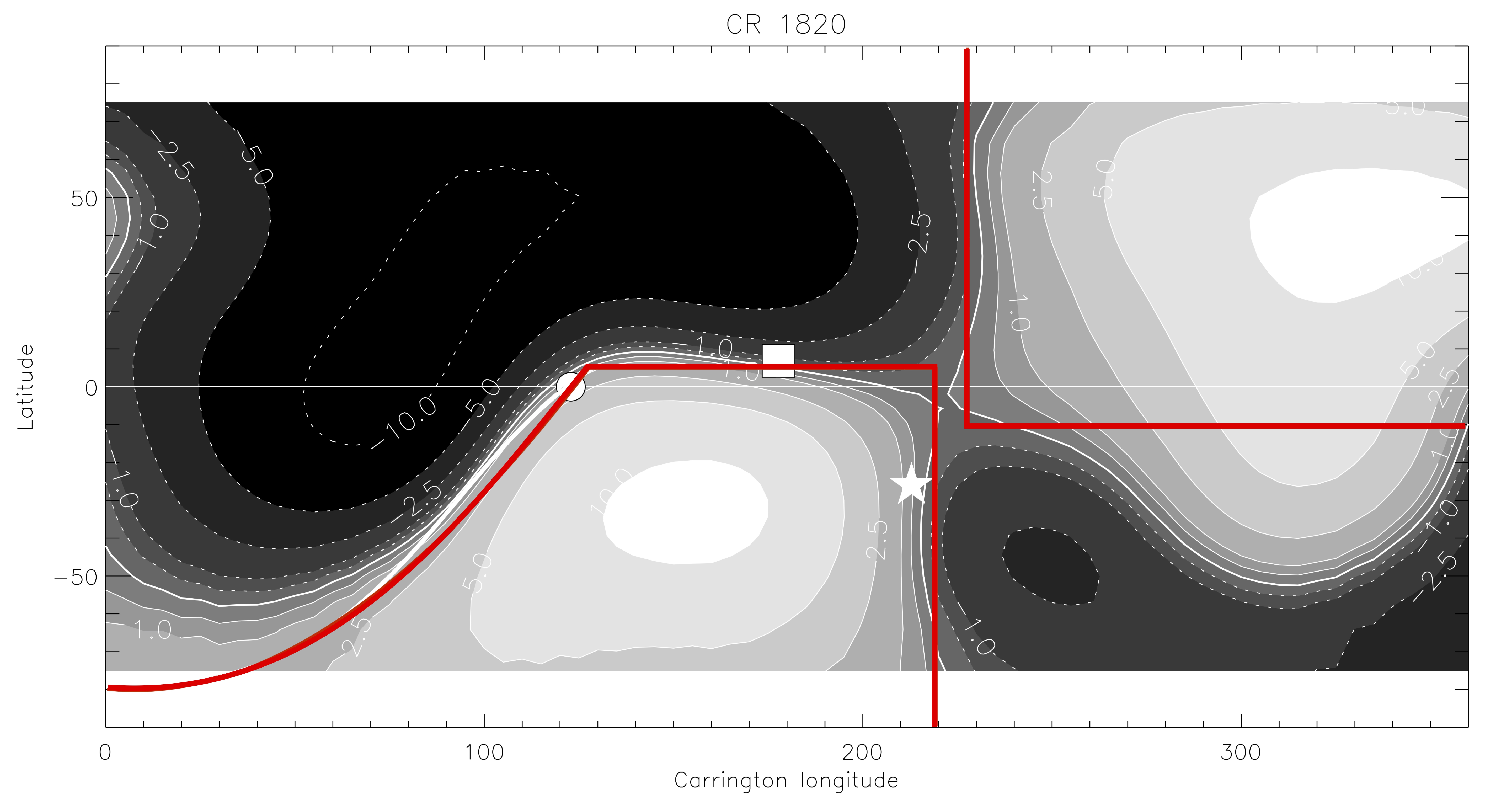}
\caption{Source surface map for Carrington rotation 1820 (1989-Sep-11 to 1989-Oct-08). The source region and Earth's relative footpoint for GLE 42 are illustrated as the star and square symbol respectively. The circle is the position of the central meridian at the time of the flare. The fit to the neutral line of the SSM is given by the red lines.} 
\label{fig.29sepssm}
\end{figure*}

The 3rd largest GLE event (since neutron monitor records began in the 1950s) is GLE 42. This event had the flare and Earth's magnetic footpoint located close to the current sheet, as is seen in the source surface map in Figure \ref{fig.29sepssm}. In contrast to GLE 69, the flare location was not in a favourable position for magnetic connectivity, occurring on the limb at S26W90 (or in some catalogs, behind, e.g. \citet{belov2010ground}). Five other GLEs have occurred behind the limb since the 1970s, all with flares within close proximity to the HCS. The flare for GLE 42 was extremely large, peaking at X9.0 \citep{miroshnichenko2000large}. There is scarce CME data associated with this event. As is seen in Figure \ref{fig.29sepssm} the current sheet during this rotation was highly disturbed and we have approximated it as an A$-$ configuration. The fit to this HCS is shown in red and is the best one achievable with the current model. The longitudinal separation between the source region and Earth's footpoint in this simulation is considerable, at nearly 40$^{\circ}$ (and 30$^{\circ}$ latitudinally). 

\begin{figure*}[t]
\epsscale{1.0}
\centering
\plotone{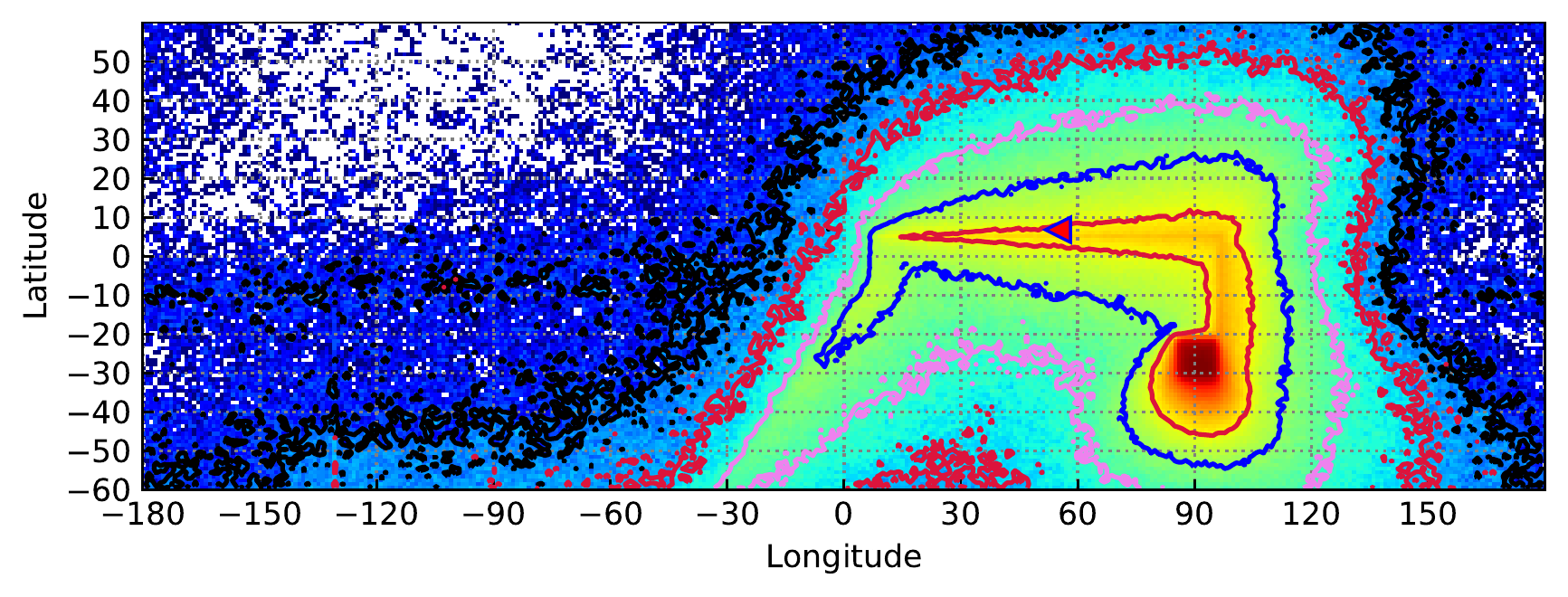}
\caption{Map plot of cumulative energetic proton crossings at 1AU for the GLE 42 simulation. The total simulation period is 72 hours. The injection region is approximated as the flare location around [90,$-$26] with the Earth's approximate location at the time of the flare shown as the red triangle. The contours illustrate the number of particle crossings, with more particle crossings focused around the source region and nearby current sheet. Corotation is included in all map plots. } 
\label{fig.29sep1au}
\end{figure*}

Figure \ref{fig.29sep1au} shows a longitude-latitude map of the cumulative proton crossings at 1AU over the total simulation period (72 hours). The parameters of the simulation, including mean free path and energy range, are described in Section \ref{sec.simulations}. In all our map plots we have included the effects of corotation. As explored by \citet{battarbee2018modeling}, corotation causes an increased longitudinal spread of particles towards the West (right). Including corotation in our results more accurately depicts what an observer at 1AU may detect. The flare is located very close to the current sheet in this instance at [90,$-$26], resulting in a concentration of increased crossings on either side of the current sheet, with a slight dip in crossings through this injection region as protons are transported away along the HCS. As this event occurred at a time of A$-$ polarity, there is drift towards the poles as well as westerly due to corotation. The main drift of the protons is towards the East along the HCS for this configuration, with the Earth (illustrated as the red triangle) favourably located in this direction and on the HCS. The darker orange region extending north and to the left from the injection location follows the shape of the current sheet and represents the large number of proton crossings in this region. The proximity of the source location to the HCS allows for this direct and efficient transport of high energy protons throughout the heliosphere. Solar wind speed observations were not available for this event, however we do not expect the approximation used here to significantly affect the results. For higher solar wind speeds, the Earth's footpoint would shift to the left but due to the shape of the HCS at that latitude the Earth would still remain in a high crossing region. 

\begin{figure*}[t]
\epsscale{0.9}
\centering
\plotone{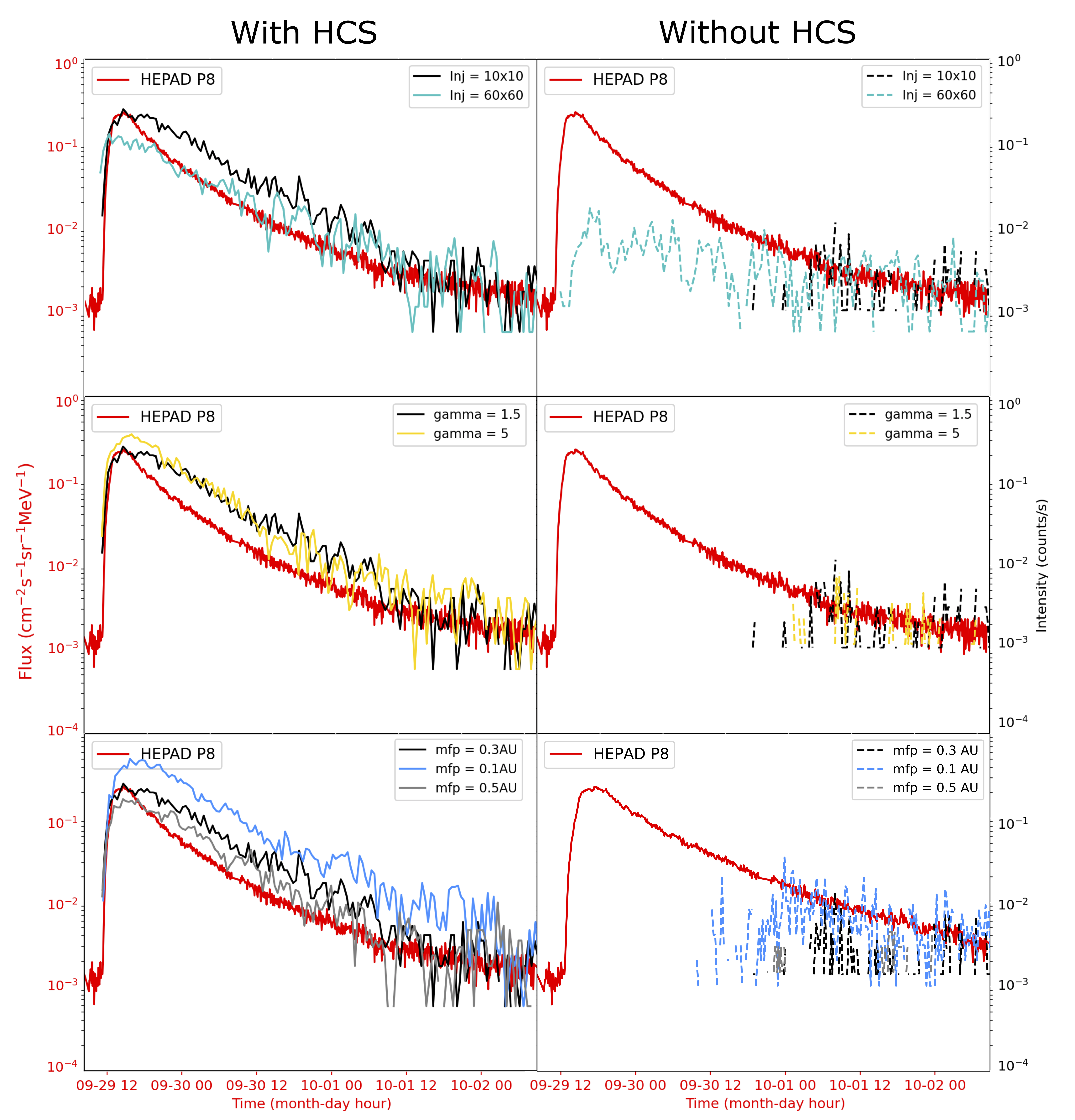}
\caption{Simulated intensity profiles at Earth's location for the GLE 42 simulations and the observed GLE 42 intensity profile for GOES-06 HEPAD (red). The top, middle and bottom row panels show the effect of varying the injection width, $\gamma$ value at injection and mean free path respectively. (Where $\gamma$ is the spectral index of the power law in energy.) All model profiles have a 10$^{\circ}$ injection width, excluding the green profile in the top row panel for a 60$^{\circ}$ injection.} 
\label{fig.29sepcompare}
\end{figure*}

Our simulations show that the particles reach Earth because of the proximity of the Earth's footpoint and the source location to the HCS. The simulated intensity profiles at Earth under a variety of parameters are shown in Figure \ref{fig.29sepcompare}. Our `standard' model (shown in black) uses an injection width of 10x10$^{\circ}$, mean free path of 0.3AU and power law index of $\gamma$\,=\,1.5. The top, middle and bottom row panels correspond to varying the injection width, power law index and mean free path respectively. The left three panels in Figure \ref{fig.29sepcompare} show these simulations with the HCS included, right three panels are without. The observed GOES-6 HEPAD profile in the P8 (355-435\,MeV for GOES-6) energy range is shown in each plot for comparison with the simulated profile. The flux, plotted on the y-axis, has units of [cm$^{-2}$\,s$^{-1}$\,sr$^{-1}$\,MeV$^{-1}$] for the HEPAD profile and [counts s$^{-1}$] for the simulated profile. Both y-axes span 4 orders of magnitude.

From Figure \ref{fig.29sepcompare} it is evident that transport along the HCS is the dominant mechanism that affects the propagation towards Earth. Whilst altering the other parameters, e.g. the injection width, causes some changes in the intensity profile, they all produce reasonable fits to the observed profile, provided the HCS is included. Even varying the power law index (central row) to $\gamma$=5 returns a similar intensity profile to $\gamma$=1.5 when the HCS is present. These simulated profiles reflect how necessary the HCS is for this simulation, and others where the source region is located close to the sheet (see discussion next on GLE 69). 

We have performed a $\chi ^{2}$ test on the simulated profiles to see which parameter fits the observations best. The $\chi ^{2}$ statistic was evaluated for the initial 24 hour time period as well as the full time range. In both cases, the `standard' model parameters (i.e. the black curve in the left hand plots) was found to achieve the best fit to the observed profile. None of the simulations without the HCS provide a reasonable similarity with the observed profiles. Similar conclusions have been found in the higher energy channels (P9 and P10) but are not shown here. 

In general, it is worth noting that our simulated and observed profiles (for HCS included) agree reasonably well. The peak flux and event duration are in good agreement for the 10$^{\circ}$ injection. The different injection widths return similar profiles, with a smaller peak flux for a larger injection width (keeping the number of particles the same). 

\subsection{GLE 69: 20 January 2005}
\label{subsec.nohcs}
\begin{figure*}
\epsscale{0.9}
\centering
\plotone{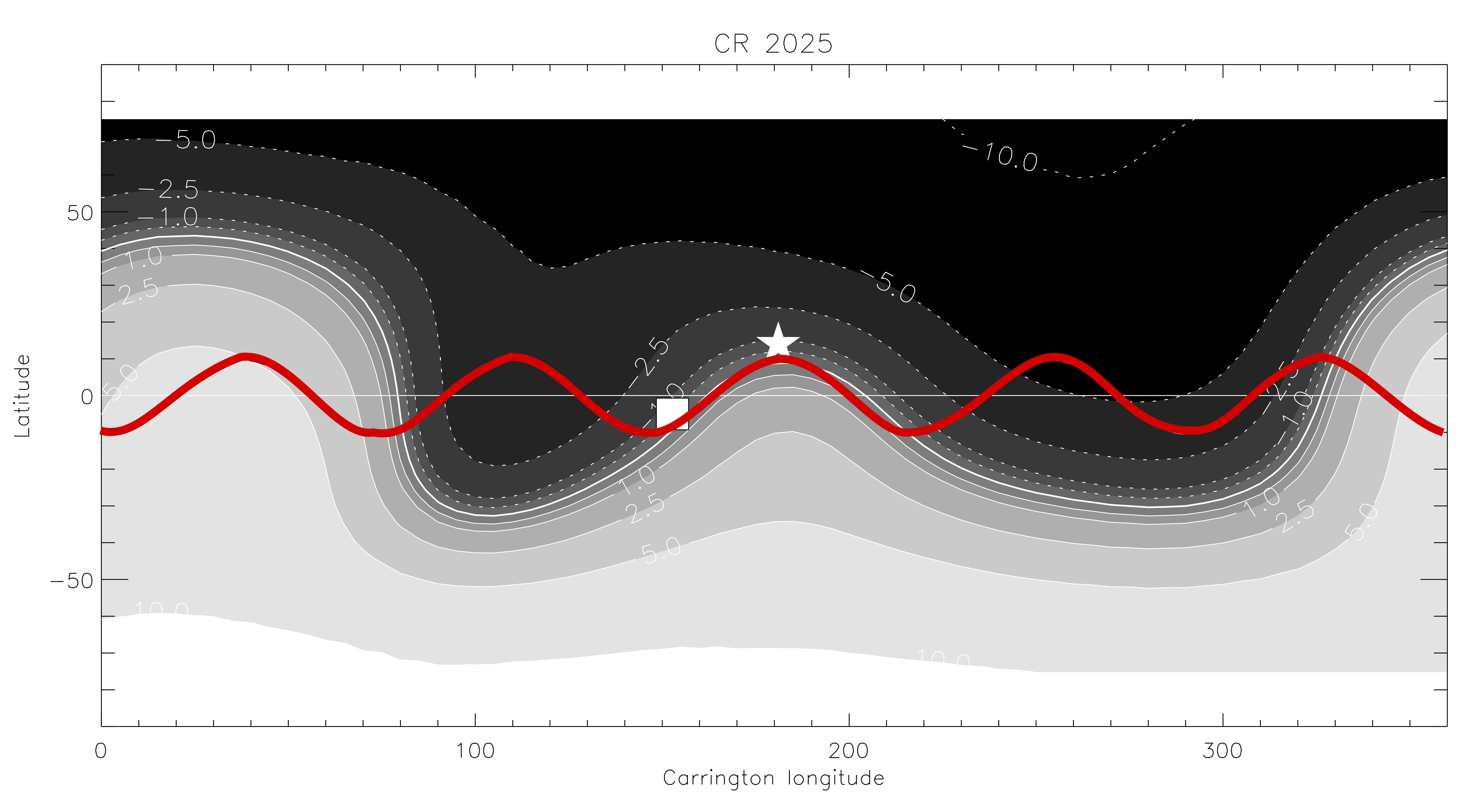}
\caption{Source surface map for Carrington rotation 2025 (2005-Jan-02 to 2005-Jan-29). The flare and Earth's relative footpoint for GLE 69 are illustrated as the star and square symbol respectively. The circle is the position of the central meridian at the time of the flare.} 
\label{fig.20janssm}
\end{figure*}

GLE 69 is one of the most intense GLEs recorded.  It it therefore of particular interest here to investigate whether the current sheet could have influenced the size of this event. Figure \ref{fig.20janssm} shows the source surface map for GLE 69. The injection location, approximated as the source location of the associated flare, is positioned at N14W63.  This western longitude of 63$^{\circ}$ is already a favourable location, given the magnetic connectivity between the Sun and Earth through the Parker spiral.  The flare, of class X7.1, was also associated with a fast halo CME travelling at speeds around 2500 km\,s$^{-1}$ (\citet{grechnev2008extreme}).  However, there are previous flares and GLEs with one or all of these conditions (large flare, fast CME, favourable position, fast solar wind speed) but much lower percentage increases in neutron monitor data. A factor that is different in GLE 69 compared to other events is the location of the flare and Earth's footpoint relative to both each other and the current sheet. From the source surface map in Figure \ref{fig.20janssm} it is clear both the flare (star) and Earth (square) are located nearly directly on the current sheet. They have a longitudinal separation of approximately 25$^{\circ}$. Despite the A$-$ configuration, the proximity of the flare to the current sheet allows for a large amount of particle drift along it as well as some towards the poles. We were able to produce a good fit to the current sheet close to the flare's location and covering Earth's footpoint location on this source surface map for use in our test particle code. As we are primarily concerned with the drift towards the Earth during this event, the fit outside this region is less relevant and does not affect our results.

\begin{figure}[t]
\epsscale{1.0}
\centering
\plotone{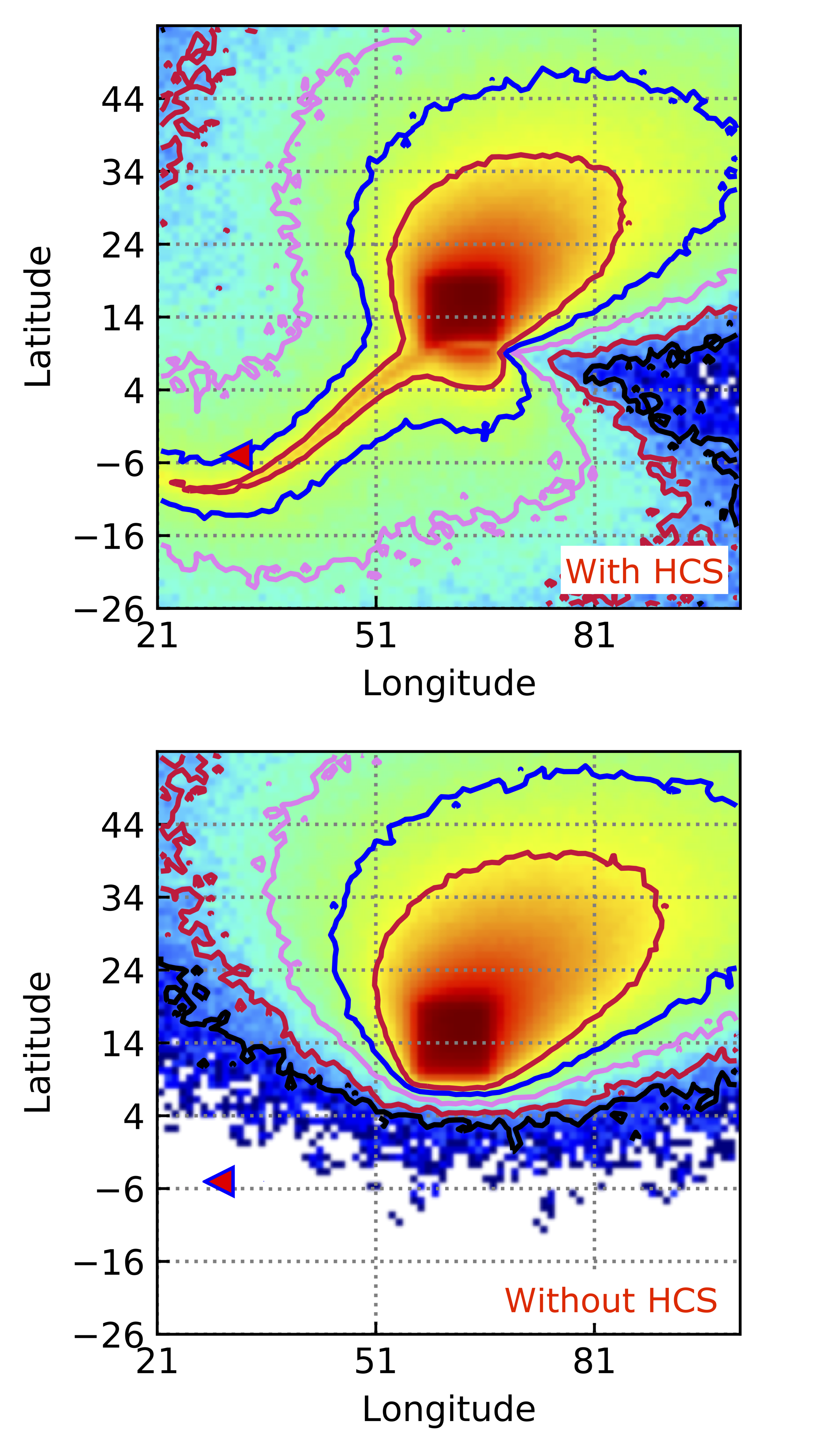}
\caption{Map plot of cumulative energetic proton crossings at 1AU for the GLE 69 simulation with the HCS (top) and without (bottom). The total simulation period is 72 hours. The injection region is approximated as the flare location around [63,14] with the Earth's approximate location at the time of the flare shown as the red triangle. The contours illustrate the number of particle crossings, with more particle crossings focused around the source region (and nearby current sheet when included).}
\label{fig.20jan1au}
\end{figure}

Figure \ref{fig.20jan1au} (top) shows the cumulative 1AU proton crossing map for the GLE 69 simulation with the HCS present. The flare is slightly displaced to the north compared to the GLE 42 simulation. The solar wind speed was increased to 820km\,s$^{-1}$ to more accurately represent the solar wind conditions during GLE 69. This solar wind speed was recorded by the CELIAS/MTOF Proton Monitor on-board the SOHO Spacecraft around the time of the flare. Similarly to Figure \ref{fig.29sep1au}, the proximity of the injection region to the HCS results in the efficient transport of protons along the current sheet towards the Earth. There is again drift towards the poles in the opposite direction, caused by the gradient and curvature drifts induced by the Parker spiral. 

\begin{figure*}[t]
\epsscale{0.9}
\centering
\plotone{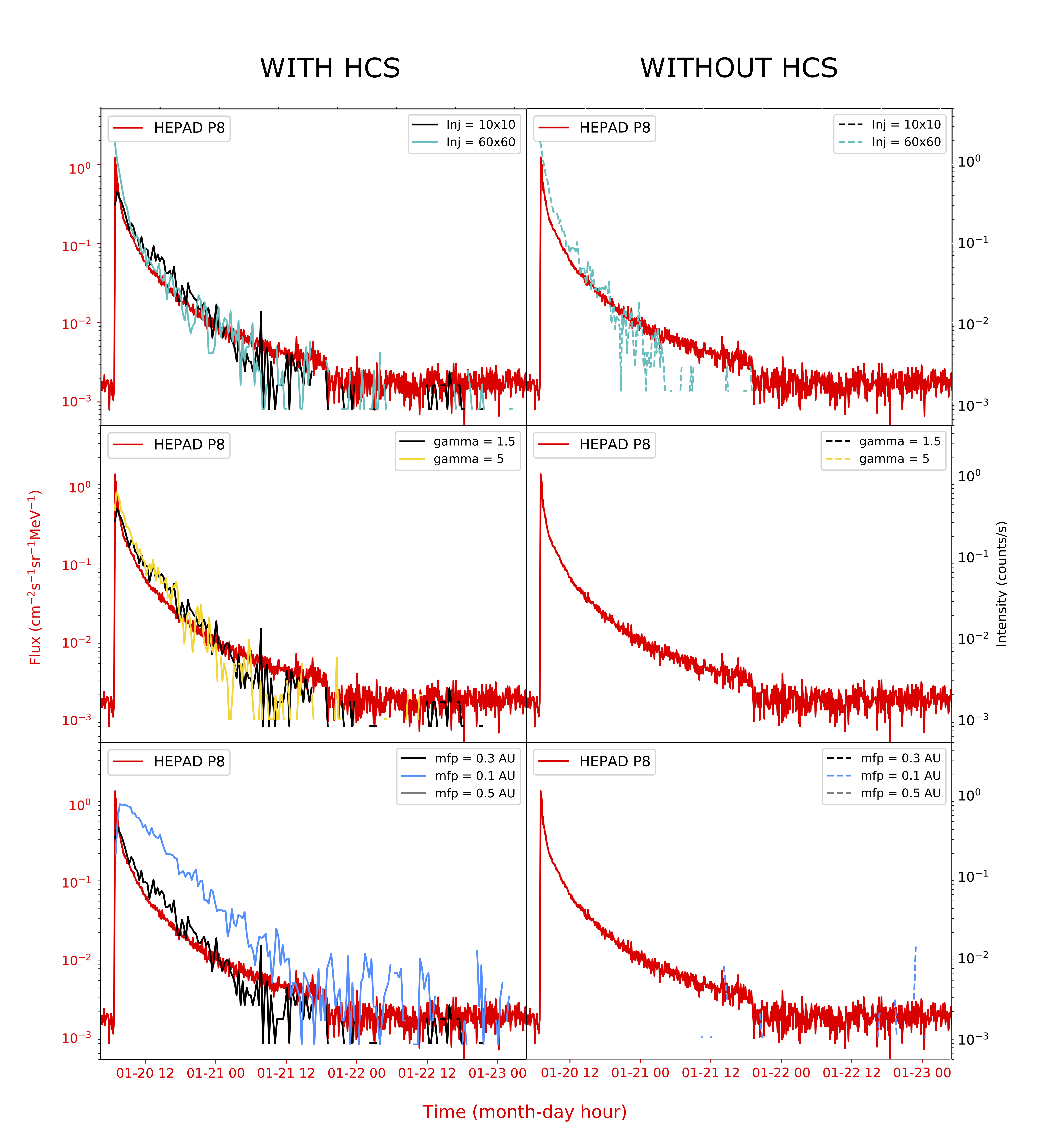}
\caption{Simulated intensity profiles at Earth's location for the GLE 69 simulations and the observed GLE 69 intensity profile for GOES-11 HEPAD (red). The top, middle and bottom row panels show the effect of varying the injection width, $\gamma$ value at injection and mean free path respectively. (Where $\gamma$ is the spectral index of the power law in energy.)  All profiles are shown for a 72 hour period.}
\label{fig.20jancompare}
\end{figure*}

The intensity profiles at Earth's footpoint at 1AU can be plotted for this simulation and are shown in Figure \ref{fig.20jancompare}. The model parameters have been varied in the same way as for Figure \ref{fig.29sepcompare} and the plot layout is the same. All profiles are plotted for approximately the same time range; over 72 hours starting 2 hours before the flare start time of 06:36 UTC (and simulation injection).

Again, the HEPAD and model profiles for this event are similar. They both peak sharply after the initial particle injection displaying an impulsive profile. Our simulations with the wavy HCS can reasonably reproduce the high energy intensity profiles observed at Earth for GLE 69 when the flare is close to the HCS. The proximity of the flare and Earth to the current sheet allows for the direct transport of high energy protons towards the Earth despite the angular separation between the two locations. 

We have calculated the $\chi ^{2}$ statistic for these events and found that the `standard' parameter values again provides the closest fit. On inspection alone it is clear that varying the injection width (top panel) has minor effects on the simulated profile and both agree well with the observations. However, removing the HCS leads to no counts at Earth for a 10$^{\circ}$ injection. There are counts for the 60$^{\circ}$ injection with no HCS and the profiles agree reasonably well early in the simulation. However, the decay is far steeper than when the HCS is included and does not fit well from around 12 hours into the simulation. 

The 1AU crossing map plots generated for GLE 69 without the HCS is shown in the bottom plot of Figure \ref{fig.20jan1au}. Earth's footpoint is again shown as the red triangle.  As can be seen in the map, no particles reach the Earth in our simulations and an intensity profile cannot be formed. This is seen in the top right hand plot of Figure \ref{fig.20jancompare}. The highest concentration of 1AU crossings are again at the injection location but the majority of the proton transport occurs towards the poles. There is also drift to the west caused by corotation. 

Comparison of the top and bottom plot in Figure \ref{fig.20jan1au} highlights how, for GLE 69, the current sheet played a pivotal role in the transport of high energy protons towards Earth. Our simulations suggest the effects observed at Earth were heavily influenced by the proximity of the injection location to the current sheet. The position of the Earth is also relevant here, as if it was located West of the flare the difference in count rates with and without the current sheet would have been less dramatic. 

Removing the current sheet for the GLE 42 simulation produces a similar result (not shown) as seen for GLE 69 in Figure \ref{fig.20jan1au} (bottom plot). The drift of the protons is focused towards the poles away from Earth's location without a HCS. As a result, no counts are produced near Earth and no comparable intensity profile can be obtained. 

The final column in Table \ref{tab.eventstab} shows which modelled events had zero counts at Earth when the HCS was removed. This was the case for 9 out of 17 events. Of the remaining 8 events, 5 of these ($^a$ in Table \ref{tab.eventstab}: GLE 52, 65-67 and 70) showed no or minimal difference in the Earth profile when the HCS was removed and thus was deemed irrelevant. The 3 remaining events ($^b$ in Table \ref{tab.eventstab}: GLE 42, 55 and 60) had minimal counts or lacked a defined profile at Earth when the HCS was removed. An improved fit (determined by visual inspection as well as $\chi ^{2}$ test result) was obtained when the HCS was included. For those cases the HCS was deemed relevant. In total, the HCS is necessary in 12 out of 17 of our simulated events in order to more closely reproduce the observed profiles at Earth.

We see sharp rises in both the observations and model results for the GLEs that have source regions close to the HCS. The HCS transport in our model appears to be a fast and efficient mechanism for this transport towards Earth. Indeed, the rise times for GLE 69 are under 30 minutes. One of the events where the inclusion of the HCS is not beneficial is GLE 65 (the first GLE in the GLE episode of late 2003). This GLE had an associated flare located much further from the HCS and is discussed next. These types of events are in the minority among GLEs. 

\subsection{October $-$ November 2003 events}
\subsubsection{GLE 65: 28 October 2003}
\label{subsec.october}
We now examine the effect of having a flare (and Earth's footpoint) located far from the current sheet. GLE 65 (28 October 2003) is the most extreme example on our list, where the flare is situated the furthest from the HCS, whilst still having a considerable neutron monitor increase of 45$\%$. 

GLE 65 is part of a multi-event episode comprised of two more GLEs, each with the flare similarly far from the sheet. The 3 GLEs (65, 66 and 67 occurring on October 28, 29 and November 2nd 2003 respectively) were all spawned from large solar eruptive events with flares of class X17, X10 and X8.3 respectively. The associated CMEs had speeds of 2459, 2029, 2598 km\,s$^{-1}$ respectively \citep{gopalswamy2012properties}. The flares and CMEs are of GLE producing caliber and are comparable to GLE 42 and 69, however the GLEs have smaller neutron monitor increases of 45$\%$, 35$\%$ and 37$\%$ respectively \citep{mccracken2012high}. The source region of GLE 65 is located at S16E08 which could explain its smaller increase at Earth, however GLE 67 is favourably located at S14W56 yet has the smallest percentage increase. The largest peak HEPAD flux was for GLE 66, followed by GLE 67 then GLE 65, however they were all of a similar peak value. The dramatic increase in both neutron monitor and HEPAD data for GLE 42 and 69 could be attributed to the current sheet's influence. Indeed, looking at the SSM for these 2003 events in Figure \ref{fig.28octSSM}, it is clear just how much further away the flare and Earth's footpoint are from the current sheet. For this A$-$ configuration the drift will be towards the northern pole with a westerly drag from corotation. While the HCS may not play a large role in these events, the Earth's footpoint is still in a favourable location to intercept the proton crossings at 1AU. As the Earth moves East relative to the flare (as seen in Figure \ref{fig.28octSSM}), it moves out of range of the particle drift and could explain why the neutron monitor increases are lower for GLE 66 and 67 compared to GLE 65.

\begin{figure*}[t]
\epsscale{0.9}
\centering
\plotone{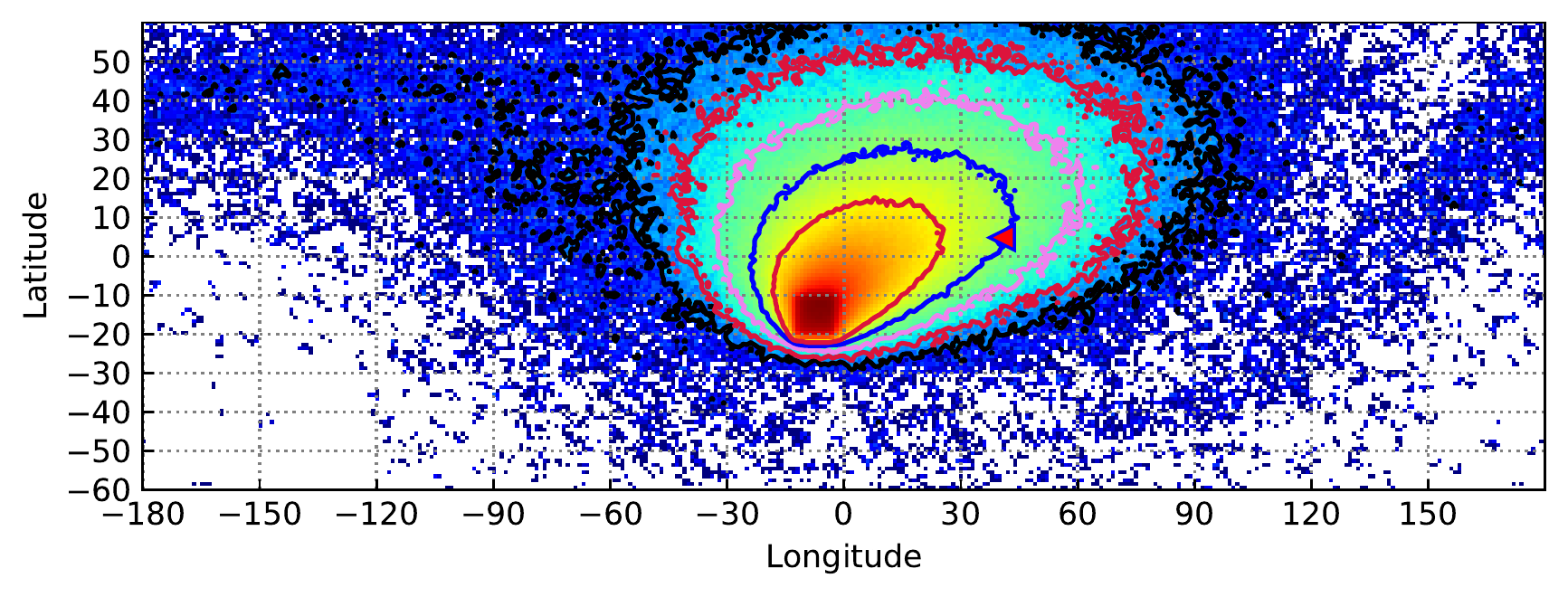}
\caption{Map plot of cumulative energetic proton crossings at 1AU for the GLE 65 simulation. The Earth's approximate location at the time of the flare shown as the red triangle. The contours illustrate the number of particle crossings, with more particle crossings focused towards the poles due to the curvature and gradient drifts. There is little drift along the HCS due to the large separation between the flare and current sheet and A$-$ polarity.} 
\label{fig.28oct03map}
\end{figure*}

The 1AU crossing map for GLE 65 is shown in Figure \ref{fig.28oct03map} across the whole longitudinal and latitudinal range. As the source location is far from the current sheet there is minimal transport along it, as is evident in this map. There are a small and diffuse number of crossings visible along it at some longitudes, however most of the particle crossings are concentrated at the injection location and away from the HCS. This simulation is a good example of where the current sheet makes a minimal difference. When the sheet is removed from the model, the profile obtained at Earth's location is virtually identical to the version with the current sheet. The generation of a GLE event for this episode can again be attributed to the favourable location of Earth's footpoint with regard to the injection location and drift direction, despite the HCS not playing a significant role.

\begin{figure*}
\epsscale{0.9}
\centering
\plotone{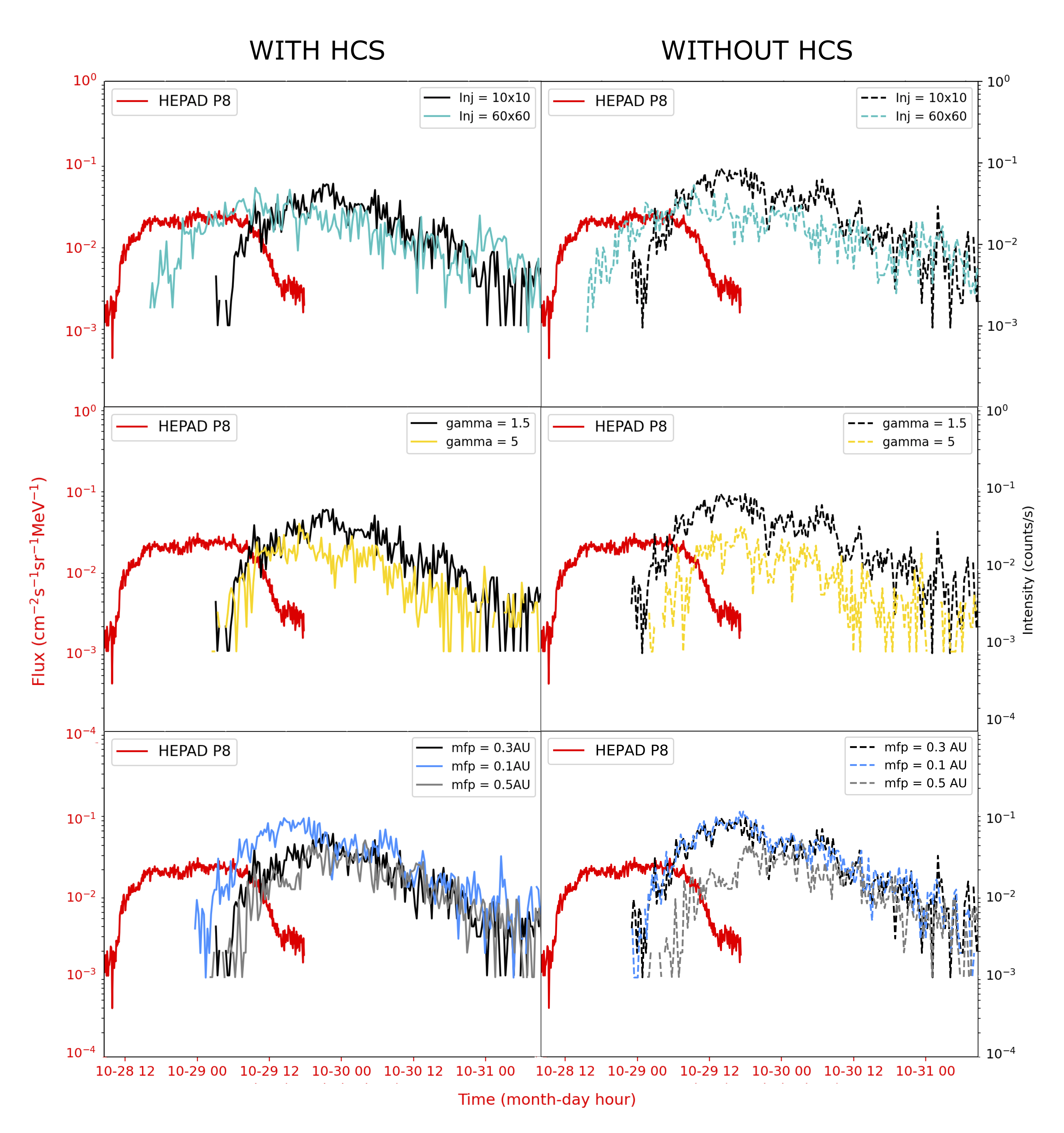}
\caption{Simulated intensity profiles at Earth's location for the GLE 65 simulations and the observed GLE 65 intensity profile for GOES-11 HEPAD (red). All profiles are shown for the P8 energy channel (350-420\,MeV). The top, middle and bottom row panels show the effect of varying the injection width, $\gamma$ value at injection and mean free path respectively. (Where $\gamma$ is the spectral index of the power law in energy.) GLE 66 occurred on 2003-Oct-29 but has been removed here to avoid confusion between profiles.}
\label{fig.28oct03compare}
\end{figure*}

The simulated intensity profiles at Earth's location for GLE 65 are shown in Figure \ref{fig.28oct03compare} with the corresponding HEPAD profile in red. These profiles are plotted for 72 hours, where the subsequent GLE (66) has been removed to avoid confusion in the plot as we only model a single injection of particles. The plots have the same layout as Figures \ref{fig.29sepcompare} and \ref{fig.20jancompare} where the same parameters are varied. 

While the intensity profiles do not agree as well as for the GLE 42 and 69 results, there are still some notable features. The test particle model produces a prolonged peak of the intensity profile, similar to that observed. However, the duration of the event is far longer and the peak time is far later from the time of particle injection. Increasing the injection width to 60$^{\circ}$ reduces this delay somewhat however the decay is far shallower than the observed profile. Increasing the injection width more shifts the simulated profile to earlier onset times however an extremely wide injection width ($>$ 120$^{\circ}$) is unreasonable to assume in these events. Varying the other parameters ($\gamma$ and mean free path) does not produce any notably better results than the `standard' model. 

The delay between initial injection and onset is thought to be due to the differing causes of these profiles when compared to GLE 42 and 69. In the earlier profiles the counts were produced as a result of HCS drifts, seemingly a more efficient process than the curvature and gradient drifts that produce the GLE 65 profiles. We acknowledge that the results for this model aren't a good representation of the observed profiles and the lack of turbulence-induced perpendicular transport is a limitation for this event. The inclusion of perpendicular transport in any future work is likely to affect model events like this, where the flare is far from the HCS, and potentially cause increased transport towards Earth over time. For the majority of other events, where the flare is closer to the HCS, the effect of adding perpendicular transport will depend on the interplay between this new mechanism and the other drift effects.

\subsubsection{4th November solar eruption}
\begin{figure*}[t]
\epsscale{0.9}
\centering
\plotone{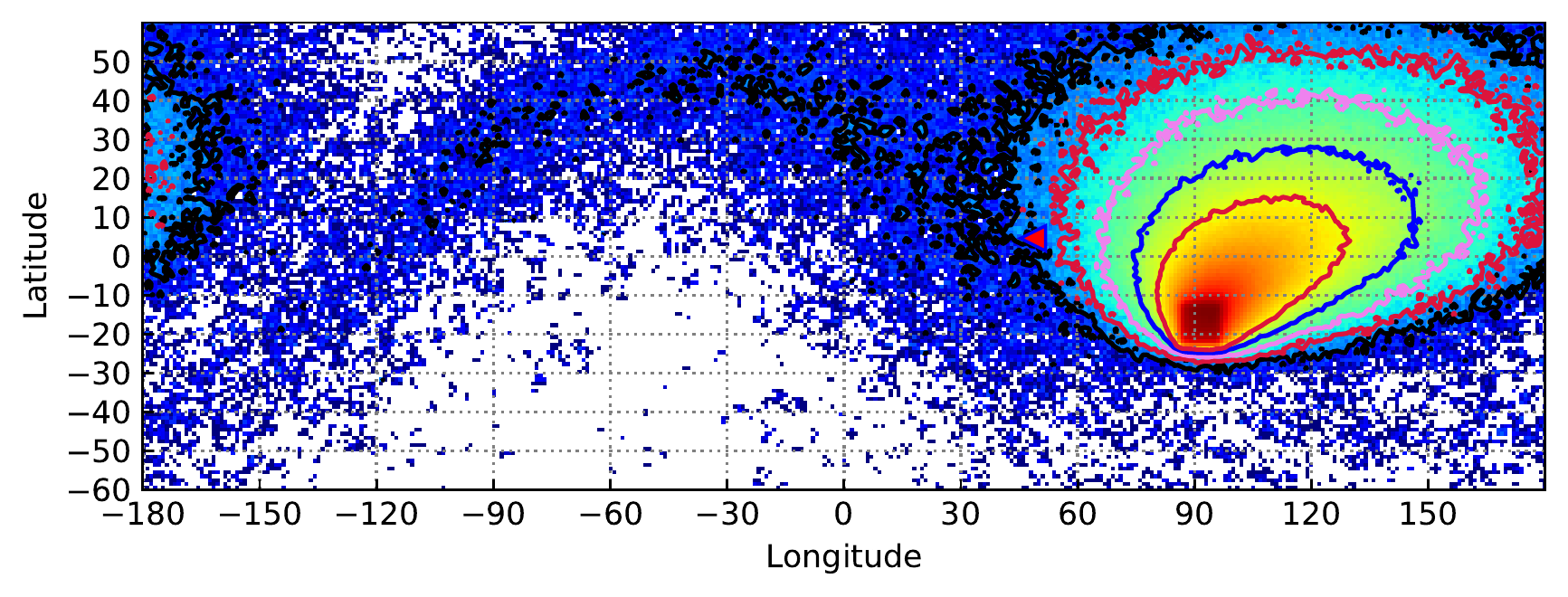}
\caption{Map plot of cumulative energetic proton crossings at 1AU for the 4 November 2003 simulation. The Earth's approximate location at the time of the flare shown as the red triangle. The contours illustrate the number of particle crossings, with more particle crossings focused towards the poles as in Figure \ref{fig.28oct03map}.} 
\label{fig.4nov03map}
\end{figure*}

The importance of the position of the Earth's footpoint relative to the injection location, even when the HCS has a minimal effect, is illustrated through modelling the 4 November 2003 event. This event was the largest flare on record, with a flare of class X28 occurring at S19W83 and associated CME velocity of 2657 km\,s$^{-1}$ \citep{gopalswamy2012properties}. It occurred from the same active region that produced GLE 65, 66 and 67, occurring two days after the latter. Given our previous results, this type of event would be expected to produce a clear event in the high energy channels at Earth. However, no significant enhancement exists in the high energy GOES HEPAD channels. From the SSM map in Figure \ref{fig.28octSSM}, the position of Earth's footpoint (labelled `4th') relative to the drift of particles from the injection location indicates that there would indeed be minimal crossings in this region as the drift direction is focused away from the Earth. The 1AU crossing map for 4 November 2003 in Figure \ref{fig.4nov03map} also illustrates this. Despite the footpoint being closer to the current sheet in Figure \ref{fig.28octSSM}, the flare is so far from the sheet there is minimal drift in that direction. For GLE 65 (which occurred 7 days prior) in Figure \ref{fig.28oct03map} the Earth is in a more favourable position to generate counts at Earth despite its flare's Easterly longitude.

Whilst the 4 November 2003 near-limb flare site might be suggested as the reason for minimal high energy flux at Earth, we have seen for GLE 42 that a behind-the-limb event is capable of producing large events down at ground-level when both the flare and Earth's footpoint are close to the current sheet. 

\section{Discussion and Conclusions} \label{sec.conclusions}
We have simulated a set of historic GLE events using a 3D test particle code to explore the role the heliospheric current sheet has on high-energy particle propagation. Our main conclusions are as follows:
\begin{itemize}
    \item Within a model that does not include perpendicular diffusion associated with turbulence, simulations of high energy SEP transport for GLE 42 (Figure \ref{fig.29sep1au}) and GLE 69 (Figure \ref{fig.20jan1au}) show that the HCS provides an important channel for propagation across the magnetic field. This result is unchanged when a larger CME-like injection of 60x60 is used. However, for GLE 69, removing the HCS still leads to some counts at Earth when a 60x60$^{\circ}$ injection is used.
    \item Additional simulations for another 15 events show that the HCS is relevant in 71$\%$ of our events. For the events where the HCS is not relevant, the source region is typically far from the HCS (e.g. GLE 65).
    \item When the source region is close to the HCS, there is significant heliospheric longitudinal transport and our model reasonably reproduces the observed high energy intensities at Earth (even when a narrow injection width of 10$^{\circ}$ is used).
    \item A study of source regions of $\geq$M7 flares demonstrated that 44$\%$ of source regions associated with GLEs were located within 10$^{\circ}$ of the HCS, while for non-GLE flare producing regions the figure is 22$\%$ (Figure ~\ref{fig.histo}).
    \item The longitudinal location of Earth relative to the source location is also relevant, with more events with larger neutron monitor increases seen when Earth is in a favourable direction from the source. This effect is also seen in the simulations, with more particle counts at Earth when it's located in the HCS drift direction.
\end{itemize}

We used our simulations to produce synthetic particle time-intensity profiles, to compare them with HEPAD observations. In addition to the three events presented in detail in Section \ref{sec.data_analysis}, we have simulated 14 other historic GLE events and found the current sheet plays a dominant role in transporting the protons towards Earth in 12 out of 17 cases. As is seen in Table \ref{tab.eventstab}, the modelled GLEs where the HCS does not aid in producing the observed profiles are GLEs 52, 65-67 and 70. For the episode containing GLE 65-67, as well as GLE 52, including the HCS makes no difference to the Earth's intensity profile because the flare is far from the HCS. For GLE 70 the separation between the flare and Earth's footpoint is $<$20$^{\circ}$ so minimal HCS transport towards Earth occurs. 

Based on the analysis of data for 8 GLEs and sub-GLEs, \citet{augusto2018relativistic} came to the conclusion that propagation along the HCS is an important mechanism that facilitates arrival of relativistic protons to Earth to produce GLEs. It is important to note that they used a different methodology compared to ours to establish whether the source active region and Earth were located within an HCS structure. They used plasma density maps from simulations from WSA-ENLIL for three events to determine the location of the active region with respect to the HCS, and 1 AU magnetic field data to gather information on Earth’s location in six events. Generally, compared to our definition their `HCS region' has a much larger extent than in our analysis. 

Our results suggest that while GLEs can occur under a variety of configurations, the spatial distribution of particles across the heliosphere is greatest when the flare site is located within 10$^{\circ}$ of the current sheet. We have also seen that the historic GLEs with the largest neutron monitor increases nearly all have flares within 10$^{\circ}$ of the HCS, suggesting that the proximity of the flare to the HCS strongly affects the severity of the event at Earth. Additionally, we have seen that more GLEs (of any magnitude) have occurred when the flare is closer to the HCS (44$\%$ of GLEs between 1976$-$2017 had a flare $<$10$^{\circ}$ from the HCS). Even for GLEs with flares occurring behind-the-limb (e.g. GLE 42, and five others since the source surface maps are first available in 1974), they all occurred within 15$^{\circ}$ of the HCS. Two of the three largest GLEs on record (69 and 42 respectively) had flares within 5$^{\circ}$ of the HCS. An additional feature of these two GLEs is the advantageous location of the Earth relative to the flare. The Earth was located East of the flare in both cases, in the direction of the particle drift (due to the A$-$ configuration). 

We have seen from GLE 65 that if the Earth's footpoint was located closer to the HCS it would generate minimal counts: it is the flare location relative to the HCS that makes the most difference in that case. It is evident the effect of the drifts due to the HCS as a requirement for a GLE cannot be unanimously ruled necessary or not for all events, and needs to be evaluated based on the configuration of the Earth and flare location relative to each other. Despite this, the majority of GLEs (for which we have source surface maps) have associated flares that occur close to the current sheet (see Figure \ref{fig.histo}) and therefore it generally needs to be considered in any future forecasting model. Currently there is not a consensus about the degree to which perpendicular diffusion effects affect SEP or GLE events nor about the relevant timescales. We can estimate the timescale for propagation of 100 MeV protons in longitude and latitude solely due to perpendicular diffusion from the results of \citet{chhiber2017cosmic} and \citet{laitinen2016solar}. We can use the perpendicular mean free path $\lambda_\perp$ in the heliospheric equatorial plane of \citet{chhiber2017cosmic} and introduce a diffusion coefficient in angular units (deg$^2$/s), given by $\kappa_\phi=\kappa_\perp/r^2 = \lambda_\perp v /(3 r^2)$ where $\kappa_\perp$ is the perpendicular diffusion coefficient, the magnetic field is assumed to be radial, $r$ is radial distance from the Sun and $v$ is the particle speed. At 1 AU \citet{chhiber2017cosmic} gives ${\lambda_\perp}$\,$\approx$\,0.006 au, resulting in ${\kappa_\phi \approx (4.5^{\circ})^2/\mathrm{hr}}$. From these considerations we can estimate that angular propagation across the field of $\Delta \phi$ = 20$^{\circ}$ via perpendicular diffusion would require a time $\tau_\perp = (\Delta \phi)^2/(2 \kappa_\phi) $ $\approx$ 9.9 hr and across 40$^{\circ}$ $\tau_\perp $ =39.5 hr. These timescales appear to be slower than those required by the HEPAD data for GLE 42 and 69.

The simulation for GLE 65 could have showed increased counts at Earth if a turbulence-induced perpendicular transport mechanism was included. Including this in our model is non-trivial and it is not included at present. There are a number of arbitrary parameters involved in building a turbulence model and little information is currently available on turbulence properties near the HCS. It is also likely that turbulence characteristics may have varied significantly in the different GLE events simulated here. Generally additional transport across the field would aid particles in reaching the HCS, where the HCS drift effects would kick in. The relative magnitude of turbulence induced cross-field motion and HCS drift will need to be clarified in future work. 


Simulations of more recent GLEs, e.g. GLE 72 illustrate these results further. For GLE 72 the flare was around 20$^{\circ}$ away from the HCS with the Earth to the East of the flare in an A+ configuration. While the fast CME (3163 km\,s$^{-1}$) and strong flare (X8.2) associated with this event likely helped cause this GLE at Earth, the neutron monitor increase was small (a maximum of around 10$\%$) and the peak P8 HEPAD flux was over an order of magnitude less than for GLE 42. If the flare had been closer to the HCS, and the Earth to the West of the flare, this flux would likely have been larger and a more powerful GLE caused. 

Some of our chosen parameter values and model properties are subject to discussion. For example, modelling broader shock regions may be necessary for some GLE events. However, injection regions as large as those for lower energy SEP events are considered unlikely \citep{nitta2012special}. The model currently includes only a single injection of particles near the Sun, whereas some episodes may include multiple events or injections in IP space. Whilst the 3D nature of the model and the inclusion of various drift effects is useful here, we hope to also include turbulence-induced perpendicular transport in the future. Indeed, some models suggest that high levels of turbulence may reduce drift effects throughout the heliosphere \citep{engelbrecht2017toward}. The interplay between drift effects and turbulence is an interesting topic worthy of further investigation. Considering gradient and curvature drifts in the Parker spiral via an analytical study that uses different models of the transport coefficients, \citet{van2021turbulent} obtained a drift suppression factor at energies of 300 MeV between 1 and 50. This factor decreases at higher energies. The suppression factors obtained do not apply to HCS drift, although \citet{van2021turbulent} commented on the possible interplay between turbulence and the HCS. Additionally, studies of GCR modulation (e.g. \cite{boschini2017solution}, \cite{song2021numerical}, \cite{aslam2021time}) also indicate that drift effects, including curvature and gradient drifts, need to be reduced at solar maximum to fit observations. The suppression of gradient and curvature drift is not as important for particles already starting close to the HCS. 

In our simulations we used a spatially constant solar wind speed based on 1 AU measurements, however due to solar wind acceleration, close to the Sun the speed is likely to be smaller than the one we considered. Given the inverse dependencies of drift velocities on solar wind speed \citep{dalla2013solar}, this will likely enhance the magnitude of gradient and curvature drift taking place close to the Sun.

The present study has focused on time intensity profiles of GLE events. However there are a number of other observables that could be used to further constrain model parameters and assess the relative contribution of drift, turbulence and other physical processes in this type of events. They include particle anisotropies and spectra. In our study we have found that the spectral index at injection does not influence intensity profiles of GLEs strongly for the events studied. However spectra may be useful to break this degeneracy and constrain the spectral index at injection.

In summary, there are a number of factors that may explain the fact that GLE events can be observed at large longitudinal and latitudinal separation from the source. Among these are: a broad injection region, efficient turbulence-induced perpendicular transport (including magnetic field line meandering) and drift effects (including HCS drift and gradient and curvature drift). With the data currently available it is difficult to isolate one mechanism as the main cause of the observed cross-field extent of GLE events, and in addition a mix of processes may be at play. Our model shows that drift effects alone can explain a large fraction of GLE events, especially those with an injection region close to the HCS, but fail to explain events like GLE 65.

We have only considered the intensity profiles at Earth's location in our simulations. Other simulated profiles can be obtained for other instruments, such as STEREO-A, STEREO-B, Messenger (e.g. similar to that done in \citet{battarbee2018multi}). Indeed, intensity profiles generated for STEREO locations could help demonstrate the proficiency in which the HCS distributes particles longitudinally through the heliosphere. 

It is easy to infer from our results that the largest events occur for situations where both the flare and Earth are located on the current sheet with the Earth in a favourable direction, however there are very few events with this configuration. Large events are very rare, and there are multiple features that combine to produce the notable results at Earth. However, our model does illustrate the necessity of the current sheet for these particular types of event. The approaching solar maximum (which will occur in an A$-$ configuration; the polarity which has provided the most intense GLEs) will hopefully produce more GLEs for study. In the mean time, modelling the full collection of largest GLEs will shed more light on the importance of the HCS during these events.

\begin{acknowledgments}
C.O.G. Waterfall and S. Dalla acknowledge support from NERC via the SWARM project, part of the SWIMMR programme (grant NE/V002864/1).

SD and TL acknowledge support from the UK Science and Technology Facilities Council (STFC) through grants ST/R000425/1 and ST/V000934/1.

This work was performed using resources provided by the Cambridge Service for Data Driven Discovery (CSD3) operated by the University of Cambridge Research Computing Service (www.csd3.cam.ac.uk), provided by Dell EMC and Intel using Tier-2 funding from the Engineering and Physical Sciences Research Council (capital grant EP/P020259/1), and DiRAC funding from the Science and Technology Facilities Council (www.dirac.ac.uk). 

SD acknowledges support from the International Space Science Institute through funding of the International Team on `Solar Extreme Events: Setting up a paradigm'.

We gratefully acknowledge the use of the SEPEM Reference Data Set version 3, European Space Agency (2021).

We acknowledge the use of data from Wilcox Solar Observatory data in this study. The Wilcox Solar Observatory is currently operated by Stanford University with funding provided by the National Science Foundation. 

We also acknowledge the use of data from the Heliophysics Event Catalogue as part of HELIO.

\end{acknowledgments}


\bibliography{bibrefs}

\begin{thebibliography}{}
\expandafter\ifx\csname natexlab\endcsname\relax\def\natexlab#1{#1}\fi
\providecommand{\url}[1]{\href{#1}{#1}}
\providecommand{\dodoi}[1]{doi:~\href{http://doi.org/#1}{\nolinkurl{#1}}}
\providecommand{\doeprint}[1]{\href{http://ascl.net/#1}{\nolinkurl{http://ascl.net/#1}}}
\providecommand{\doarXiv}[1]{\href{https://arxiv.org/abs/#1}{\nolinkurl{https://arxiv.org/abs/#1}}}

\bibitem[{Aslam {et~al.}(2021)Aslam, Bisschoff, Ngobeni, Potgieter, Munini,
  Boezio, \& Mikhailov}]{aslam2021time}
Aslam, O., Bisschoff, D., Ngobeni, M., {et~al.} 2021, The Astrophysical
  Journal, 909, 215

\bibitem[{Augusto {et~al.}(2019)Augusto, Navia, de~Oliveira, Nepomuceno, Fauth,
  Kopenkin, \& Sinzi}]{augusto2018relativistic}
Augusto, C., Navia, C., de~Oliveira, M., {et~al.} 2019, Publications of the
  Astronomical Society of the Pacific, 131, 024401

\bibitem[{Battarbee {et~al.}(2017)Battarbee, Dalla, \&
  Marsh}]{battarbee2017solar}
Battarbee, M., Dalla, S., \& Marsh, M.~S. 2017, The Astrophysical Journal, 836,
  138

\bibitem[{Battarbee {et~al.}(2018{\natexlab{a}})Battarbee, Dalla, \&
  Marsh}]{battarbee2018modeling}
---. 2018{\natexlab{a}}, The Astrophysical Journal, 854, 23

\bibitem[{Battarbee {et~al.}(2018{\natexlab{b}})Battarbee, Guo, Dalla,
  Wimmer-Schweingruber, Swalwell, \& Lawrence}]{battarbee2018multi}
Battarbee, M., Guo, J., Dalla, S., {et~al.} 2018{\natexlab{b}}, Astronomy \&
  Astrophysics, 612, A116

\bibitem[{Belov {et~al.}(2010)Belov, Eroshenko, Kryakunova, Kurt, \&
  Yanke}]{belov2010ground}
Belov, A., Eroshenko, E., Kryakunova, O., Kurt, V., \& Yanke, V. 2010,
  Geomagnetism and Aeronomy, 50, 21

\bibitem[{Bieber {et~al.}(2004)Bieber, Evenson, Dr{\"o}ge, Pyle, Ruffolo,
  Rujiwarodom, Tooprakai, \& Khumlumlert}]{bieber2004spaceship}
Bieber, J.~W., Evenson, P., Dr{\"o}ge, W., {et~al.} 2004, The Astrophysical
  Journal Letters, 601, L103

\bibitem[{Bieber {et~al.}(2002)Bieber, Droege, Evenson, Pyle, Ruffolo, Pinsook,
  Tooprakai, Rujiwarodom, Khumlumlert, \& Krucker}]{bieber2002energetic}
Bieber, J.~W., Droege, W., Evenson, P.~A., {et~al.} 2002, The Astrophysical
  Journal, 567, 622

\bibitem[{Boschini {et~al.}(2017)Boschini, Della~Torre, Gervasi, Grandi,
  J{\'o}hannesson, Kachelriess, La~Vacca, Masi, Moskalenko, Orlando,
  {et~al.}}]{boschini2017solution}
Boschini, M., Della~Torre, S., Gervasi, M., {et~al.} 2017, The Astrophysical
  Journal, 840, 115

\bibitem[{Chhiber {et~al.}(2017)Chhiber, Subedi, Usmanov, Matthaeus, Ruffolo,
  Goldstein, \& Parashar}]{chhiber2017cosmic}
Chhiber, R., Subedi, P., Usmanov, A.~V., {et~al.} 2017, The Astrophysical
  Journal Supplement Series, 230, 21

\bibitem[{Cliver {et~al.}(2020)Cliver, Mekhaldi, \&
  Muscheler}]{cliver2020solar}
Cliver, E., Mekhaldi, F., \& Muscheler, R. 2020, The Astrophysical Journal
  Letters, 900, L11

\bibitem[{Dalla \& Browning(2005)}]{dalla2005particle}
Dalla, S., \& Browning, P. 2005, Astronomy \& Astrophysics, 436, 1103

\bibitem[{Dalla {et~al.}(2020)Dalla, De~Nolfo, Bruno, Giacalone, Laitinen,
  Thomas, Battarbee, \& Marsh}]{dalla20203d}
Dalla, S., De~Nolfo, G., Bruno, A., {et~al.} 2020, Astronomy \& Astrophysics,
  639, A105

\bibitem[{Dalla {et~al.}(2013)Dalla, Marsh, Kelly, \&
  Laitinen}]{dalla2013solar}
Dalla, S., Marsh, M., Kelly, J., \& Laitinen, T. 2013, Journal of Geophysical
  Research: Space Physics, 118, 5979

\bibitem[{Dalla {et~al.}(2015)Dalla, Marsh, \& Laitinen}]{dalla2015drift}
Dalla, S., Marsh, M., \& Laitinen, T. 2015, The Astrophysical Journal, 808, 62

\bibitem[{Engelbrecht {et~al.}(2017)Engelbrecht, Strauss, Le~Roux, \&
  Burger}]{engelbrecht2017toward}
Engelbrecht, N., Strauss, R., Le~Roux, J., \& Burger, R. 2017, The
  Astrophysical Journal, 841, 107

\bibitem[{Forbush(1946)}]{forbush1946three}
Forbush, S.~E. 1946, Physical Review, 70, 771

\bibitem[{Gopalswamy {et~al.}(2013)Gopalswamy, Xie, Akiyama, Yashiro, Usoskin,
  \& Davila}]{gopalswamy2013first}
Gopalswamy, N., Xie, H., Akiyama, S., {et~al.} 2013, The Astrophysical Journal
  Letters, 765, L30

\bibitem[{Gopalswamy {et~al.}(2012)Gopalswamy, Xie, Yashiro, Akiyama,
  M{\"a}kel{\"a}, \& Usoskin}]{gopalswamy2012properties}
Gopalswamy, N., Xie, H., Yashiro, S., {et~al.} 2012, Space Science Reviews,
  171, 23

\bibitem[{Grechnev {et~al.}(2008)Grechnev, Kurt, Chertok, Uralov, Nakajima,
  Altyntsev, Belov, Yushkov, Kuznetsov, Kashapova,
  {et~al.}}]{grechnev2008extreme}
Grechnev, V., Kurt, V., Chertok, I., {et~al.} 2008, Solar Physics, 252, 149

\bibitem[{Hu {et~al.}(2017)Hu, Li, Ao, Zank, \&
  Verkhoglyadova}]{hu2017modeling}
Hu, J., Li, G., Ao, X., Zank, G.~P., \& Verkhoglyadova, O. 2017, Journal of
  Geophysical Research: Space Physics, 122, 10

\bibitem[{Kahler(1994)}]{kahler1994injection}
Kahler, S. 1994, The Astrophysical Journal, 428, 837

\bibitem[{Kelly {et~al.}(2012)Kelly, Dalla, \& Laitinen}]{kelly2012cross}
Kelly, J., Dalla, S., \& Laitinen, T. 2012, The Astrophysical Journal, 750, 47

\bibitem[{Kocharov {et~al.}(2020)Kocharov, Pesce-Rollins, Laitinen, Mishev,
  K{\"u}hl, Klassen, Jin, Omodei, Longo, Webb,
  {et~al.}}]{kocharov2020interplanetary}
Kocharov, L., Pesce-Rollins, M., Laitinen, T., {et~al.} 2020, The Astrophysical
  Journal, 890, 13

\bibitem[{Kubo {et~al.}(2009)Kubo, Nagatsuma, \& Akioka}]{kubo20092}
Kubo, Y., Nagatsuma, T., \& Akioka, M. 2009, Journal of the National Institute
  of Information and Communications Technology Vol, 56

\bibitem[{K{\"u}hl {et~al.}(2017)K{\"u}hl, Dresing, Heber, \&
  Klassen}]{kuhl2017solar}
K{\"u}hl, P., Dresing, N., Heber, B., \& Klassen, A. 2017, Solar Physics, 292,
  1

\bibitem[{Laitinen {et~al.}(2018)Laitinen, Effenberger, Kopp, \&
  Dalla}]{laitinen2018effect}
Laitinen, T., Effenberger, F., Kopp, A., \& Dalla, S. 2018, Journal of Space
  Weather and Space Climate, 8, A13

\bibitem[{Laitinen {et~al.}(2016)Laitinen, Kopp, Effenberger, Dalla, \&
  Marsh}]{laitinen2016solar}
Laitinen, T., Kopp, A., Effenberger, F., Dalla, S., \& Marsh, M. 2016,
  Astronomy \& Astrophysics, 591, A18

\bibitem[{McCracken {et~al.}(2012)McCracken, Moraal, \&
  Shea}]{mccracken2012high}
McCracken, K., Moraal, H., \& Shea, M. 2012, The Astrophysical Journal, 761,
  101

\bibitem[{McCracken {et~al.}(2008)McCracken, Moraal, \&
  Stoker}]{mccracken2008investigation}
McCracken, K., Moraal, H., \& Stoker, P. 2008, Journal of Geophysical Research:
  Space Physics, 113

\bibitem[{Miroshnichenko {et~al.}(2000)Miroshnichenko, De~Koning, \&
  Perez-Enriquez}]{miroshnichenko2000large}
Miroshnichenko, L., De~Koning, C., \& Perez-Enriquez, R. 2000, Space Science
  Reviews, 91, 615

\bibitem[{Nitta {et~al.}(2012)Nitta, Liu, DeRosa, \&
  Nightingale}]{nitta2012special}
Nitta, N., Liu, Y., DeRosa, M., \& Nightingale, R. 2012, Space science reviews,
  171, 61

\bibitem[{Onsager {et~al.}(1996)Onsager, Grubb, Kunches, Matheson, Speich,
  Zwickl, \& Sauer}]{onsager1996operational}
Onsager, T., Grubb, R., Kunches, J., {et~al.} 1996, in GOES-8 and Beyond, Vol.
  2812, International Society for Optics and Photonics, 281--290

\bibitem[{Papaioannou {et~al.}(2016)Papaioannou, Sandberg, Anastasiadis,
  Kouloumvakos, Georgoulis, Tziotziou, Tsiropoula, Jiggens, \&
  Hilgers}]{papaioannou2016solar}
Papaioannou, A., Sandberg, I., Anastasiadis, A., {et~al.} 2016, Journal of
  Space Weather and Space Climate, 6, A42

\bibitem[{Reames(2009)}]{reames2009solar}
Reames, D.~V. 2009, The Astrophysical Journal, 693, 812

\bibitem[{S{\'a}iz {et~al.}(2005)S{\'a}iz, Ruffolo, Rujiwarodom, Bieber, Clem,
  Evenson, Pyle, Duldig, \& Humble}]{saiz2005relativistic}
S{\'a}iz, A., Ruffolo, D., Rujiwarodom, M., {et~al.} 2005, in 29th
  International Cosmic Ray Conference (ICRC29), Volume 1, Vol.~1, 229

\bibitem[{Schwenn(2006)}]{schwenn2006space}
Schwenn, R. 2006, Living Reviews in Solar Physics, 3, 1

\bibitem[{Shea \& Smart(2012)}]{shea2012space}
Shea, M., \& Smart, D. 2012, Space science reviews, 171, 161

\bibitem[{Siscoe(2000)}]{siscoe2000space}
Siscoe, G. 2000, Journal of Atmospheric and Solar-Terrestrial Physics, 62, 1223

\bibitem[{Song {et~al.}(2021)Song, Luo, Potgieter, Liu, \&
  Geng}]{song2021numerical}
Song, X., Luo, X., Potgieter, M.~S., Liu, X., \& Geng, Z. 2021, The
  Astrophysical Journal Supplement Series, 257, 48

\bibitem[{Tooprakai {et~al.}(2016)Tooprakai, Seripienlert, Ruffolo, Chuychai,
  \& Matthaeus}]{tooprakai2016simulations}
Tooprakai, P., Seripienlert, A., Ruffolo, D., Chuychai, P., \& Matthaeus, W.
  2016, The Astrophysical Journal, 831, 195

\bibitem[{Usoskin {et~al.}(2020)Usoskin, Koldobskiy, Kovaltsov, Gil, Usoskina,
  Willamo, \& Ibragimov}]{usoskin2020revised}
Usoskin, I., Koldobskiy, S., Kovaltsov, G., {et~al.} 2020, Astronomy \&
  Astrophysics, 640, A17

\bibitem[{van~den Berg {et~al.}(2021)van~den Berg, Engelbrecht, Wijsen, \&
  Strauss}]{van2021turbulent}
van~den Berg, J., Engelbrecht, N., Wijsen, N., \& Strauss, R. 2021, The
  Astrophysical Journal, 922, 200

\bibitem[{Winterhalter {et~al.}(1994)Winterhalter, Smith, Burton, Murphy, \&
  McComas}]{winterhalter1994heliospheric}
Winterhalter, D., Smith, E., Burton, M., Murphy, N., \& McComas, D. 1994,
  Journal of Geophysical Research: Space Physics, 99, 6667

\end{thebibliography}
\bibliographystyle{aasjournal}



\end{document}